\documentclass[review]{elsarticle}

\usepackage{lineno,hyperref,graphicx,adjustbox,subcaption}
\modulolinenumbers[5]

\journal{Nuclear Inst. and Methods in Physics Research, A}

\begin{document}

\begin{frontmatter}

\title{The STAR MAPS-based PiXeL detector}

\author[a]{Giacomo Contin\corref{mycorrespondingauthor}}
\cortext[mycorrespondingauthor]{Corresponding author}
\ead{giacomo.contin@gmail.com}
\author[a]{Leo Greiner}
\author[b]{Joachim Schambach}
\author[c]{Michal Szelezniak}
\author[a]{Eric Anderssen}
\author[a]{Jacque Bell}
\author[a,+]{Mario Cepeda}
\author[a]{Thomas Johnson}
\author[a,d]{Hao Qiu}
\author[a]{Hans-Georg Ritter}
\author[a]{Joseph Silber}
\author[a]{Thorsten Stezelberger}
\author[a,e]{Xiangming Sun}
\author[a]{Co Tran}
\author[a]{Chinh Vu}
\author[a]{Howard Wieman}
\author[a]{Kenneth Wilson}
\author[a]{Rhonda Witharm}
\author[a]{Samuel Woodmansee}
\author[a]{John Wolf}

\address[a]{Lawrence Berkeley National Laboratory, 1 Cyclotron Road, Berkeley, CA 94720, USA}
\address[b]{University of Texas, 1 University Station, Austin, TX 78712, USA}
\address[c]{Institut Pluridisciplinaire Hubert Curien (IPHC), Strasbourg, France}
\address[d]{now at Purdue University, West Lafayette, IN 47907, USA}
\address[e]{now at Central China Normal University, Wuhan, Hubei 430079, China}

\address[+]{deceased}




\begin{abstract}
The PiXeL detector (PXL) for the Heavy Flavor Tracker (HFT) of the STAR experiment at RHIC is the first application of the state-of-the-art thin Monolithic Active Pixel Sensors (MAPS) technology in a collider environment. Custom built pixel sensors, their readout electronics and the detector mechanical structure are described in detail. Selected detector design aspects and production steps are presented. The detector operations during the three years of data taking (2014-2016) and the overall performance exceeding the design specifications are discussed in the conclusive sections of this paper.
\end{abstract}


\end{frontmatter}

\tableofcontents
\section{Introduction}\label{sec:introduction}
We describe the pixel detector system (PXL) for the STAR experiment at the Relativistic Heavy Ion Collider (RHIC) at the Brookhaven National Laboratory. The STAR PXL detector is the first large-scale application of the state-of-the-art thin Monolithic Active Pixel Sensors (MAPS) technology in a collider environment. PXL is a part of a 3-detector system called Heavy Flavor Tracker (HFT) that has been added to the pre-existing STAR apparatus just before the 2014 RHIC Run to significantly improve the impact parameter resolution of STAR tracking and to enable the direct topological reconstruction of hadronic decays of heavy flavor mesons and baryons in the heavy ion collision environment. After introducing the HFT physics motivations in this section, the paper describes the PXL detector design requirements in Section~\ref{sec:requirements} and gives an overview of the HFT system in Section~\ref{sec:overview}. The detector characteristics are discussed in detail in the following sections, focusing on the MAPS sensor (Section~\ref{sec:sensor}), electronics (Section~\ref{sec:electronics}), mechanics and cooling (Section~\ref{sec:mechanics}) respectively. Section~\ref{sec:production} describes the PXL detector production process and Section~\ref{sec:operations} summarizes the detector integration and operations during the three years of data taking (2014-2016). The detector performance measured in the 2014 Run data is finally shown in Section~\ref{sec:performance}. Selected lessons learned from the PXL project are summarized in Section~\ref{sec:lessons}. The conclusions and an outlook on future particle physics applications of the MAPS technology are presented in Section \ref{sec:conclusions}.
\subsection{Physics motivations}\label{sec:physicsmot}
One of the main goals of the STAR experiment at RHIC is to study p+p, p+Au, d+Au, and Au+Au collisions at several energies up to $\sqrt{s_{NN}}$~=~200~GeV for A+A and up to 500 GeV for p+p collisions with the aim to reproduce and characterize the QCD phase transition between hadrons and partons \cite{bib1}. Heavy quark measurements are a key component of the heavy ion program for the systematic characterization of the dense medium created in heavy ion collisions, the so-called Quark-Gluon Plasma (QGP). Due to their mass, heavy quarks are only produced by hard processes early in the collision and not by thermal processes after the equilibration of the plasma, which makes mesons containing heavy quarks (e.g. charm, \emph{c}) an ideal probe for studying the initial conditions of the produced QGP. The main tracking detector used in the STAR experiment is a Time Projection Chamber (TPC), with $|\eta|\leq$~1 and full azimuthal coverage, operated inside a 0.5 T magnetic field. With its 1~mm pointing resolution, the TPC is not able to resolve the decay vertices of short-lived particles, like D$^0$(c$\bar u$) mesons ($c\tau$~$\sim$~120~$\mu$m) and $\Lambda_c$(udc) baryons ($c\tau$~$\sim$~60~$\mu$m), from the collision primary vertex. Introducing the HFT detector inside the TPC Inner Field Cage significantly improves the system track pointing resolution. The HFT and its 4 layers of silicon detectors with graded spatial resolution enable tracking inwards from the TPC and achieving a track pointing resolution of the order of 30~$\mu$m for 1~GeV/c or larger momentum particles at the vertex.

\subsection{HFT system overview}
The Heavy Flavor Tracker (HFT) in STAR consists of 4 concentric cylindrical silicon detector layers with three different sensor technologies \cite{TDR}. 
The outermost layer at 22 cm radius from the beam line is called the Silicon Strip Detector (SSD) \cite{SSD}; it is based on double-sided silicon strip sensors with 95~$\mu$m inter-strip pitch and 35 mrad relative P- N-side stereo angle inclination. The SSD silicon and front-end chips were part of an existing past STAR detector and have been equipped with new, faster readout electronics to match the increased readout speed ($>$~1~kHz) of the upgraded STAR experiment. The SSD consists of 20 ladders, each with 16 sensors for a total ladder length of 106 cm. The total number of channels in the SSD is approximately 5$\cdot$10$^5$. The detector is air-cooled, which allows for a low radiation length of approximately 1\% $X_0$.
Inside the SSD, the Intermediate Silicon Tracker (IST) layer is placed at a radius of 14~cm. It is based on single sided silicon pad sensors with a 600~$\mu$m~$\times$~6~mm pitch. The IST is composed of 24 ladders, each equipped with 6 silicon pad sensors and a readout chip, for a total sensitive ladder length of 50~cm. The total number of channels in the IST is just above 1.1$\cdot$10$^5$. The IST is liquid cooled with aluminum cooling tubes integrated into the ladder structure, which results in a total material budget smaller than 1.5\% radiation length.
The two innermost layers at 8 and 2.8 cm radii constitute the PiXeL (PXL) detector, based on state-of-the-art CMOS Monolithic Active Pixel Sensors (MAPS). A total of 400 MAPS sensors are distributed over 40 ladders (10 at the inner PXL layer radius and 30 at the outer radius) and cover a surface area of 0.16~m$^2$ with 356~M square pixels. A more detailed description of the PXL detector is provided later in this paper.
Equipped with this new micro-vertex detector, STAR is able to provide a distance of closest approach (DCA) pointing resolution of less than 50~$\mu$m for 750~MeV/c kaons, which enables the topological reconstruction of decay vertices of heavy flavor particles, like D$^0$ mesons ($c\tau$~$\sim$~120~$\mu$m), in the high-multiplicity environment typically produced in Au-Au collisions at 200~GeV. This resolution is achieved by tracking inwards from the TPC, which provides a pointing resolution of approximately 1~mm, through the SSD and IST, with pointing resolutions of 250-300~$\mu$m, to the PXL detector, that can point at secondary vertices with the resolution of a few tens of micrometers. 

\section{PXL detector design requirements and choices}\label{sec:requirements}

\subsection{Detector requirements}
The PXL detector has been designed in order to achieve the physics goals described in Section \ref{sec:physicsmot}. The track pointing resolution is primarily determined by the two innermost measurements of the track position. This resolution is improved by placing the first detection layer as close to the beam line as possible, minimizing the material budget to reduce the multiple-scattering track distortion for low-momentum tracks, and selecting the sensor segmentation that maximizes the intrinsic single-layer spatial resolution of the reconstructed track points. Fine segmentation and a short integration time window are needed to minimize the event pile-up and keep the detector occupancy low.
The PXL design is also constrained by the existing STAR detector layout and environment. The PXL is designed to match the TPC acceptance in $\eta$ and $\varphi$, and the beam pipe radius (20~mm) provides a mechanical limit for the minimum radius of the innermost PXL layer. The detector has to survive the radiation level expected in STAR.
On the basis of these considerations and of extensive simulations, the PXL Detector has been designed in order to meet the following requirements:
\begin{itemize}
\item $|\eta|\leq$~1 and full azimuthal coverage
\item DCA pointing resolution $\leq$~60~$\mu$m required for 750 MeV/c kaons
\begin{itemize}
\item Two or more layers with a separation of $>$~5~cm
\item Pixel size of $\leq$~30~$\mu$m
\item Radiation length $\leq$~0.5\% per layer,  including support structure, with 0.37\% per layer as goal
\end{itemize}
\item Integration time of $<$~200~$\mu$s
\item Sensor efficiency $\geq$~99\% with accidental rate $\leq$~10$^{-5}$
\item Radiation tolerance up to 90~kRad/year and 2$\cdot$10$^{11}$ to 10$^{12}$~1MeV~n$_{eq}$/cm$^2$
\end{itemize}

\subsection{Technology choices}
The technology and architecture have been chosen in order to meet these requirements and are reflected in our detector design. These design choices include:
\begin{itemize}
\item MAPS  technology, providing low power dissipation and short integration time
\item thinned sensors and low-mass cable with low radiation length
\item air-cooling, to minimize the material budget
\item support mechanics designed for quick detector installation or replacement
\item pixel positions fully mapped
\end{itemize}

The implementation of these choices in the PXL design is described in the next section.

\section{System overview}\label{sec:overview}

\subsection{Global layout}
\begin{figure}[h!]
\centering 
\includegraphics[width=.85\textwidth]{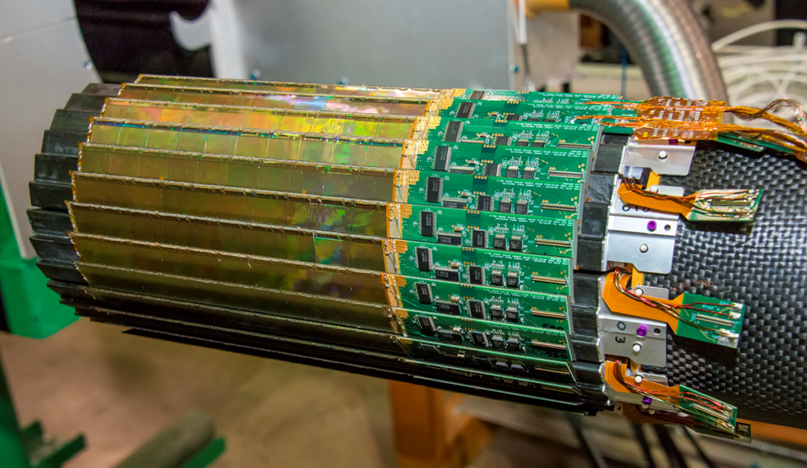}
\caption{\label{fig:PXLpic}The HFT PXL detector.}
\end{figure}

The PXL detector, shown in Figure \ref{fig:PXLpic}, consists of two cylindrical layers of CMOS Monolithic Active Pixel Sensors (MAPS) located at radii of 2.8 and 8 cm. The total of 400 MAPS sensors covers the surface area of 0.16~m$^2$ with 356M pixels and pixel pitch of 20.7~$\mu$m. 

\begin{figure}[h!]
\centering 
\includegraphics[width=.85\textwidth]{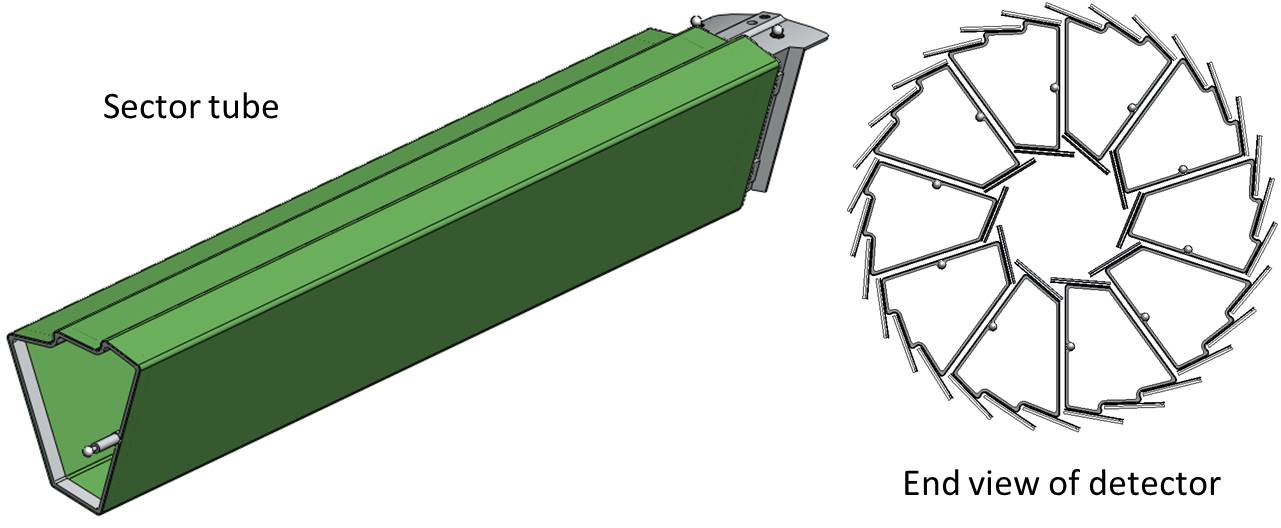}
\caption{\label{fig:PXL_sector_tube}Sector tube shape and end view of the detector showing the ladder positions on the sector.}
\end{figure}

Mechanically, the PXL detector is subdivided into two detector-halves attached on one end to a set of unique cantilevered mechanics, which allows for the fast insertion and retraction of the detector while preserving the pixel positional stability at the level of 20~$\mu$m, as described in detail in Section \ref{sec:mechanics}. The same mechanical support serves as air delivery and extraction ducts for the detector air cooling system.
The PXL detector has been designed as a highly parallel system \cite{Leo2011}, where each half consists of 5 sectors mounted to a detector half shell using precision machined mounts. A sector represents the basic unit in terms of powering and readout. Each sector consists of a trapezoidal, thin (250~$\mu$m) carbon fiber sector tube supporting four 10-sensor ladders, one at the inner radius, and three at the outer radius, arranged in a turbo geometry design (see Figure \ref{fig:PXL_sector_tube}).

Each sector is composed of 40 MAPS sensors and is serviced by a single Mass Termination Board (MTB), where all differential and single-ended signals are buffered and where the ladder power supplies are regulated (see Section \ref{sec:MTB} for details). The MTBs are connected with sectors using 2~m long cables and are attached to the detector mechanical support (5 boards on each detector-half structure). 
Further downstream, in the readout electronics rack, the data stream of each sector (two data outputs per sensor give 80 LDVS pairs) is handled by a custom built FPGA-based readout board. All ten readout boards that service the complete PXL detector operate in parallel and are triggered by the STAR trigger system. All of the PXL sensors are readout continuously and the readout boards process their data streams to form event-based data blocks in response to the STAR triggers. The event-formed PXL data are then sent from readout boards to DAQ PCs for integration with global STAR event data structures.

Details of the PXL electronics system structure and implementation are presented in Section \ref{sec:electronics}.

\section{The MAPS \emph{Ultimate-2} sensor}\label{sec:sensor}

\subsection{Sensor development}
The silicon sensors used in the PXL detector are Monolithic Active Pixel Sensors or MAPS, implemented in a standard commercially available CMOS technology. They integrate both sensor and readout electronics in one silicon device \cite{FirstMAPS}. This technology provides the best compromise between several design parameters critical for the PXL detector. MAPS devices combine excellent position resolution with sufficient radiation tolerance and low power dissipation. Limited power dissipation allows these sensors to be operated at room temperature with air-cooling that is optimal for minimizing the material budget. The MAPS production is highly cost effective compared to alternative technologies such as hybrid pixels, as it relies on standard CMOS processes and the device thinning to 50~$\mu$m is a routine operation performed by standard industry services. 

 The MAPS sensor that is used in the PXL detector represents the third and final sensor generation developed by the PICSEL group at Institut Pluridisciplinaire Hubert Curien (IPHC), Strasbourg, France, specifically for this detector. The sensor development started in 2003 with early, small-size sensors of 128$\times$128 pixels with analog readout and a signal integration time of 4~ms, later scaled into a larger prototype of 320$\times$640 pixels. The second generation of full-reticule size sensors (640$\times$640 pixels) featured a sequential, binary readout of all pixels with integration time of 640~$\mu$s. The third and final generation, described below in detail, includes on-chip data sparsification and integration time of 186~$\mu$s. The architectural advances between the three sensor generations benefited from the continuous MAPS development at IPHC spanning different application domains. The close collaboration between LBNL and IPHC allowed for an optimized and synchronized development of the sensors and the detector readout system.

\subsection{Sensor design}
The \emph{Ultimate-2} sensor (also known as Mimosa28) developed for the HFT application is a Monolithic Active Pixel Sensor (MAPS)  fabricated in the 0.35~$\mu$m, twin-well technology, with a high-resistivity ($\geq$~400~$\Omega\cdot$cm) epitaxial layer (15~$\mu$m) for optimized signal-to-noise performance and radiation hardness \cite{RadHard}. The active volume of the sensor extends to approximately the thickness of the epitaxial layer and the charge collecting diode, N-well/P-epi, is biased by an internal voltage of approximately 0.8~V and the grounded substrate. The Ultimate-2 sensor uses the same rolling-shutter readout architecture and digital processing techniques as Mimosa26, but has been optimized for radiation tolerance, readout speed and active surface area to match the PXL detector requirements \cite{MIMOSA4HFT}. The main features of the Ultimate-2 sensor are listed in Table \ref{tab:sensorspecs}. A detailed description of this sensor and its internal structure that extends beyond the scope of this section is available in \cite{Valin,Ultimate2}.

\begin{table}[h!]
\begin{center}
  \begin{tabular}{|  l  |  p{4.8cm} | }
 \hline   
    & \emph{Ultimate-2} \\
\hline
    Pixel Size & 20.7~$\mu$m~$\times$~20.7~$\mu$m \\
Array size & 928~$\times$~960 \\
Frame integration time = readout time & 185.6~$\mu$s \\
Noise after CDS & 10-12 e$^-$ \\
S/N & $\sim$30 \\
Collected charge (MIP) & $\sim$1000~e$^-$ \\
Data output & 2 LVDS lines \\
Data sparsification & $\leq$~9 1D-clusters per row, \\ 
& $\leq$~4 pixels per 1D-cluster \\
Configuration protocol  & JTAG \\
\hline

    \end{tabular}
    \caption{\emph{Ultimate-2} sensor main characteristics.}
    \label{tab:sensorspecs}
    \end{center}
\end{table}

\begin{figure}[htbp]
\centering 
\includegraphics[width=.42\textwidth]{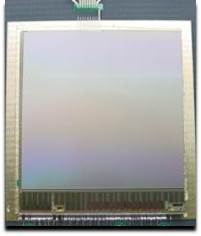}
\qquad
\includegraphics[width=.5\textwidth]{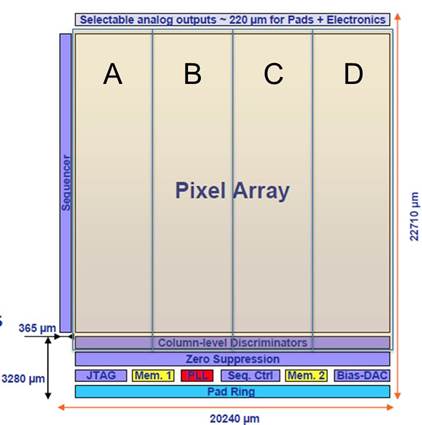}
\caption{\label{fig:Ultimate2}Left: Picture of an Ultimate-2 sensor wire-bonded to a testing board. The bonding pads at the top are connected to the analog outputs and were used only for testing purposes during the development phase. Right: Block-diagram of the Ultimate-2 sensor.}
\end{figure}

The Ultimate-2 sensor is an array of 928~$\times$~960 square pixels (a total of $\sim$~890k pixels on a 20.22~$\times$~22.71~mm$^2$ die) with a pitch of 20.7~$\mu$m (see Figure \ref{fig:Ultimate2}. Each pixel includes circuitry for readout, amplification, and Correlated Double Sampling (CDS) for signal extraction and noise subtraction. The readout of the pixels is performed for all columns in parallel, one row at a time, through discriminators with programmable thresholds at the end of each column \cite{Discriminators}. The digitized signals are then passed through zero-suppression circuitry located below the pixel array on the same chip. The sensor data format employs run-length encoding on the row data which delivers an encoded hit addresses combined with the number of adjacent hits to that address in the cluster for up to 9 hit clusters per row to on-chip memory for intermediate buffering.

The pixel size was chosen to provide good spatial resolution and to balance power consumption and radiation hardness, which have opposing requirements on the number of pixels per surface area. The pixel array is divided into four sub-arrays to allow compensation of process variations by independent configuration of reference voltages for each sub-array. The pixel array is read out one row at a time in a rolling-shutter fashion and the full array processing time defines the sensor integration time, which is equal to 185.6~$\mu$s. The zero suppression system processes one row of data at a time and encodes up to 9 hit clusters per row. The encoded data are then stored in an on-chip memory for intermediate buffering. The memory is structured as two banks of 1800 words each that allow simultaneous reads and writes to this memory. One bank will be used for writing to it, while the second bank can be read at the same time. At the end of the readout cycle the roles of the memory banks are exchanged, i.e. the bank that was previously written to is now available for reading, while the previously read from bank can now be written to. The data are read out via two Low-Voltage Differential Signaling (LVDS) outputs per sensor with a clock speed of 160~MHz. The sensor operating at 3.3~V dissipates approximately 700~mW distributed between the analog pixel array, 27\%, and the analog and digital periphery, 42\% and 31\% respectively. This power budget translates into approximately 150~mW/cm$^{2}$ which allows for operation at room temperature with simple air cooling. Configuration of the operational parameters of the sensors, including discriminator threshold, as well as programming of several test modes, is achieved over five signal lines using the built-in Joint Test Action Group (JTAG) protocol. 

This sensor achieves an ionizing particle detection efficiency of more than ~99\% with an accidental noise hit rate of $<$~10$^{-5}$ even after exposure to an ionizing radiation dose of 150 kRad and non-ionizing radiation dose of 2$\cdot$10$^{12}$ 1MeV~n$_{eq}$/cm$^2$. The single hit spatial resolution of this sensor is 6~$\mu$m (binary resolution) but can be improved to 3.7~$\mu$m through cluster centroid reconstruction methods \cite{Valin}.

Several characteristics of the PXL detector sensor were built into the chip to improve its testability and optimize the detector construction process. Notably, different test modes can be configured using the JTAG interface to verify the functionality of specific circuit sections independently of the others, including pixel readout before zero suppression and the readout of programmable pixel patterns to validate the functionality of the digital processing circuitry. The chip carries four dedicated fiducial marks that were used for mapping sensor locations in the PXL detector system. All bonding pads, each 120~$\times$~80~$\mu$m$^2$ in size, are located along one side of the sensor and the pad ring includes two pads per I/O of the sensor to provide separate sets of pads for probe testing and for wire-bonding. This doubling of I/O pads preserves the high quality of the metal surface on the main pads for wire bonding.

\subsection{Sensor testing and characterization}\label{sec:sensorQA}

Assembling a detector composed of ladders with 50~$\mu$m thick sensors poses unique challenges in terms of handling and testing of silicon devices.  To assure high yield in ladder construction it was decided to probe test sensors as the last step before gluing them onto ladders, i.e. after thinning and dicing. The construction of the PXL detector included building two full copies of the detector and a set of spare components, equivalent to a third detector. 

\begin{figure}[h!]
\centering 
\includegraphics[width=.85\textwidth]{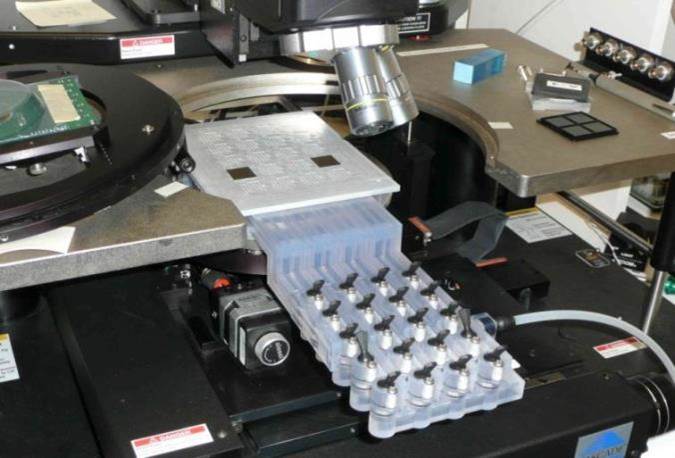}
\caption{\label{fig:probechuck}The probe testing system setup with the custom built aluminum vacuum chuck.}
\end{figure}

This necessitated fabricating $\sim$3600 full production sensors assuming reasonable fabrication and thinning yields at the level of 60\% and 90\%, respectively. The wafers had an overall mechanical thinning and dicing yield of about 93\%. Only the mechanically intact sensors have been tested and contributed to the functional yields discussed in this section. To perform this task efficiently, we developed an automated probe-testing system using a standard automated probe station that helped us assess the quality of sensors based on their electrical performance, including the identification of malfunctioning pixels. 
The probe card used in our system is a printed circuit board equipped with 49 test probe needles accompanied by two sets of probe touch sensors installed to detect mechanical contact at the edges of the die. In addition, the card hosted signal buffers and voltage regulators to provide full sensor readout at the nominal clock frequency of 160~MHz.
The main factors contributing to the test automation and efficiency were a custom built vacuum chuck capable of holding up to 20 thinned silicon dies at the same time and a set of scripted testing procedures.
The chuck (shown in Figure \ref{fig:probechuck} in the probe testing system setup) has been built using a 3D printed rapid prototyping structure glued to an aluminum plate. The plastic element provided a set of vacuum channels with individual on/off valves that allowed the operator to load the 50~$\mu$m thin sensors one by one and keep them flat while loading the others. The aluminum plate provided a high precision surface with chip alignment ridges for each die to facilitate precise manual positioning. Once all the sensors were aligned and their positions registered in the probe station control software developed in the LabWindows/CVI environment, the test system automatically stepped through each position and executed a set of scripted test procedures without human intervention. Test procedures included the functional verification of the configuration registers, scanning of reference voltage settings (e.g. signal detection thresholds) and extraction of noise performance parameters for both the temporal noise and fixed pattern noise. A typical test cycle would contain 18 sensors and last approximately 6 hours.

\subsection{Test parameters and yields}
The test parameters have been broken into two groups: functional parameters and performance parameters. Sensors that fail one of the functional parameter tests do not qualify for use in the detector. Sensors that do pass all the functional parameter tests are graded based upon a number of performance criteria.

The set of functional parameters includes:
\begin{itemize}
\item current consumption
\item optimized reference voltage
\item clamping voltage
\item JTAG functionality
\item High-speed readout functionality
\end{itemize}

The set of performance parameters includes: 
\begin{itemize}
\item Number of malfunctioning\footnote{A pixel is defined as malfunctioning if not sensitive to discriminator threshold variations.} pixels
\item Thermal noise and fixed pattern noise
\end{itemize}

Sensors with thermal noise and fixed pattern noise deviating by less than 4~$\sigma$ from the average and with no malfunctioning pixels, are graded as top quality sensors (\emph{Tier-1}). Sensors with the same characteristics but showing less than 0.05\% malfunctioning pixels are accepted as \emph{Tier-2} sensors. \emph{Tier-1} and \emph{Tier-2} sensors were used for the construction of the detectors and the spare ladders for refurbishment. These requirements guarantee good performance uniformity and an overall detector active fraction larger than 99.9\% also after refurbishment. The overall yield of the detector grade sensors for the different batches tested during the PXL detector production varied between 46\% and 60\%, mainly depending on the deterioration of the probe card pins in use. A close monitoring of the yield and a periodic probe card replacement and repair, and the repetition of false negative tests, allowed for an overall detector grade yield of $\sim$55\%, with $\sim$45\% and $\sim$10\% for \emph{Tier-1} and \emph{Tier-2} respectively. Eventually, sensors with a fraction of malfunctioning pixels between 0.05\% and 0.5\% (graded as \emph{Tier-3}) were also used for the construction of additional spare ladders available for test purposes and as surplus stock. The yield for \emph{Tier-3} was $\sim$10\%. About 86\% of the rejected sensors showed an operating current differing from the nominal value by more than 15\%. The remainder was rejected for higher number of dead pixels and for noise level out of specifications. 


\section{Electronics}\label{sec:electronics}

\subsection{System architecture}
The readout electronics are divided into 10 identical parallel systems following the mechanical segmentation of the PXL detector into sectors \cite{JS}. At the end of each ladder and out of the low-mass region are readout buffers and cable drivers that send the binary zero-suppressed data over 2~m of custom low-mass twisted pair cable (42 AWG, 63~$\mu$m diameter) to the ``Mass Termination Board'' (MTB).  These cables are characterized by low stiffness and significantly reduce possible distortions into the mechanical structure. The MTB provides additional buffering and drives the signals to the Readout (RDO) board; it also provides latch-up protected power supplies for the ladders. Each MTB services 4 ladders (one sector), i.e. there are 10 MTB’s in the PXL detector. The sensor data are then transmitted over approximately 13~m of twisted-pair cable (VHDCI cable, 34 pairs, 30 AWG, 250~$\mu$m diameter) to an RDO board in the low radiation area of the STAR experimental hall. The ``Field Programmable Gate Array'' (FPGA) based RDO board receives the data, performs trigger based hit selection, buffers, and formats the resulting data into event structures, and then sends it over 100 m optical fibers to one of ten fiber readout channels mounted in two PCs in the DAQ room. These DAQ PCs are connected to the rest of STAR DAQ for event building, where the PXL data are combined with the data from the other STAR detector subsystems providing data for the same event before sending it to final storage. The complete readout system for PXL consists of 10 RDO boards, corresponding to the 10 PXL sectors, mounted in one 9U-size crate just outside the STAR detector. In addition to providing the readout of the PXL sensors and the interface to the STAR trigger and DAQ, the RDO boards also provide monitoring data to the STAR Slow Controls system and receive configuration data for the PXL sensors from a control PC in the counting house. The control PC interfaces to the STAR Slow Controls system to provide monitoring and control of the PXL system to the STAR shift crew.

\subsection{Low-mass cable}
\begin{figure}[h!]
\centering 
\includegraphics[width=.85\textwidth]{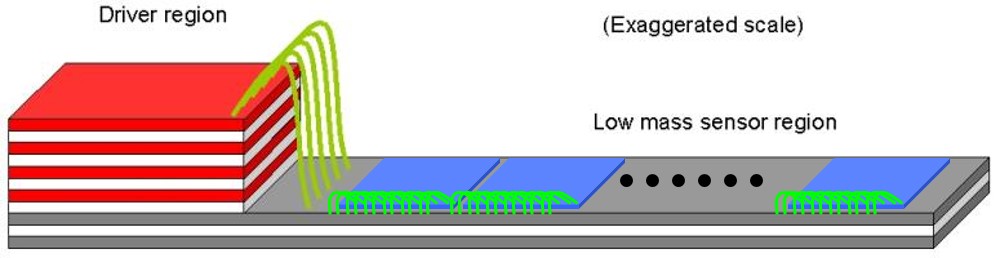}
\caption{\label{fig:ladderscheme}Exaggerated scale representation of the STAR PXL low mass cable.}
\end{figure}

In addition to providing signal and power connections to the sensors from the insertion side of the ladders, the design of the ladder FPC is an important part of meeting the material budget for the detector. This cable is located under the sensors and is designed to have two major components as seen in Figure \ref{fig:ladderscheme}. 

The primary section of the low mass cable is the two-layer aluminum conductor flex printed circuit (FPC) that is located in the low mass region of the detector acceptance. The aluminum gives the FPC a low averaged radiation length of 0.128\% including the bypass capacitors and the encapsulant that covers the wire bonds. The sensors are wire bonded to the FPC and the FPC carries the sensor electrical connections (data, clock, configuration, power, ground, etc.) out of the low mass detector acceptance region to a driver board. The driver board is a standard multi-layer PCB that is glued to the surface at the end of the aluminum conductor FPC. The electrical connections between the driver board and the FPC are made with wire bonds. The driver board provides the routing to and from driver chips that then drive the data and configuration connections to the Mass Termination Board (MTB). In this way, the sensors need only drive the data and control signals the 20cm length of the low mass region of the ladder cable and the radiation length in the region of interest is minimized.

The PXL low-mass cable is a double sided design in aluminum conductor technology with 30~$\mu$m aluminum conductor on both sides and a 50~$\mu$m kapton dielectric. The minimum size features are simple 125~$\mu$m traces and gaps with vacuum deposition plated through vias, ENIG finish and standard solder mask. These cables were fabricated in the CERN PCB shops.

\subsection{Mass Termination Board}\label{sec:MTB}
There are 10 MTB’s in the PXL detector system. Each board services one sector (four ladders).
The MTB performs three distinct functions:
\begin{enumerate}
\item provides an additional stage of signal buffering between a sector and an RDO board,
\item delivers regulated power supplies to the ladders,
\item integrates circuitry for parameter adjustment and remote monitoring.
\end{enumerate}

\begin{figure}[h!]
\centering 
\includegraphics[width=.85\textwidth]{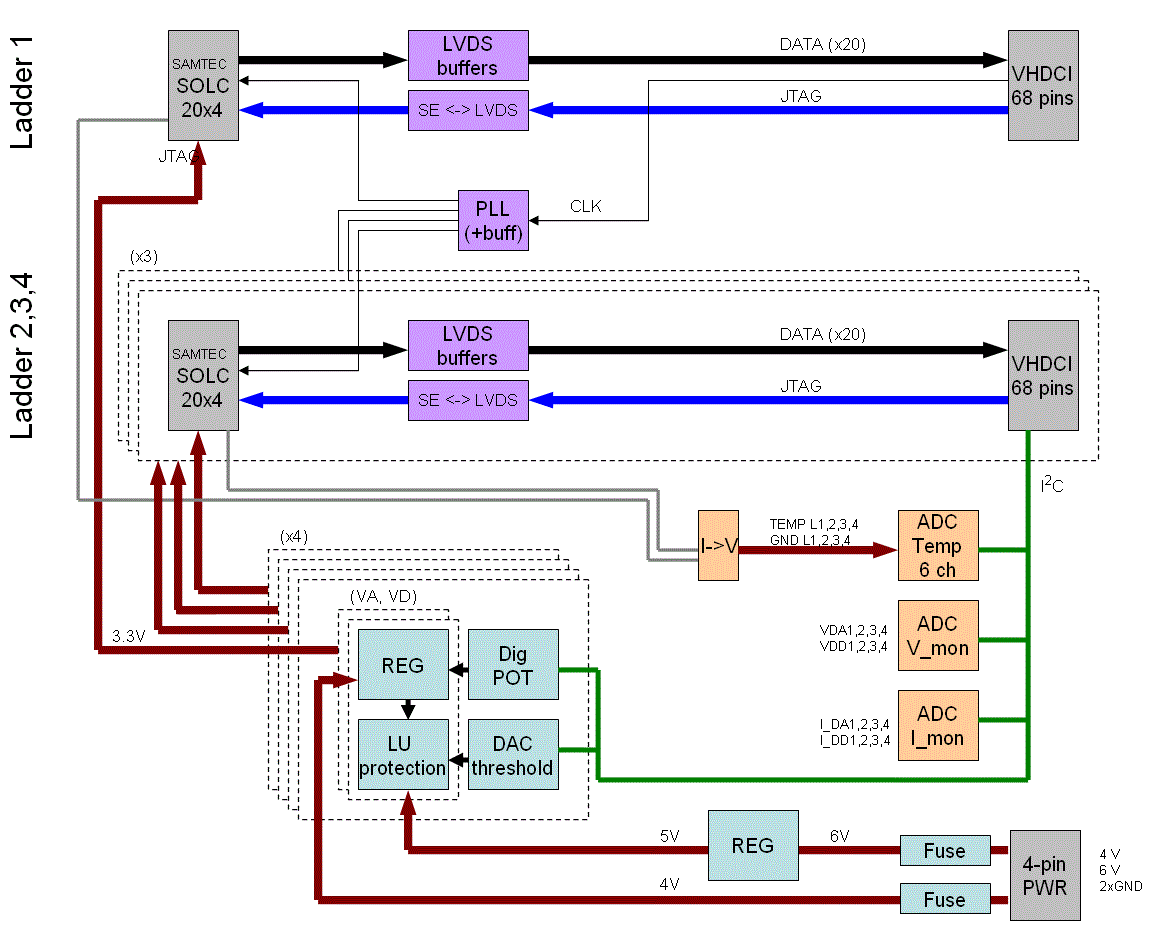}
\caption{\label{fig:MTB_blockdiagram}Block diagram of the Mass Termination Board. This board services once sector by providing over-current protected power supplies to the ladders and signal buffering between ladders and the RDO board.}
\end{figure}

A block diagram showing the functional components of the MTB is shown in Figure \ref{fig:MTB_blockdiagram}.

The signal buffering implemented on the MTB improves the signal integrity in the long connections between the RDO and ladders that is split between two different types of cables: fine-wire TP cable (2~m) and more robust VHDCI cables (13~m). In addition to LVDS buffering of data and clock lines and single-ended to differential translation of the sensor JTAG control interface, the board provides also a simple clock distribution network. A single clock signal arriving from the RDO board is fanned out and distributed to all four ladders to guarantee synchronous operation of all 40 sensors.

There are 10 voltage regulators on the MTB board. Eight of them provide analog and digital power supplies for each ladder individually, one provides a power supply to all on-board buffers, and one powers all of the over-current monitoring circuitry. The role of the over-current monitoring circuitry is to provide a protection against latch-up for all ladder power supplies and the MTB buffers. With the exception of the power supply for the current monitoring, all other voltage regulators are equipped with individual and independent over-current protection circuitry. Each circuit monitors the current consumption at the regulator output using a 50~m$\Omega$ shunt resistor and a differential amplifier with a gain of 10 that feeds into a comparator with latch function. The signal in the differential amplifier passes through a low pass filter with the corner frequency of approximately 16~Hz. The current threshold at the comparator input is configured using an ``Inter-Integrated Circuit'' (I2C) controlled DAC with a resolution of 20~mA/bit. When the measured current exceeds the threshold, a fast latching comparator disables the corresponding voltage regulator. The state of each regulator is monitored by the RDO board and a disabled regulator stays inactive until it is reset by a signal from the RDO board. This approach provides a reliable latch-up protection and is compatible with an automated latch-up recovery procedure implemented in the firmware of the RDO board.

The MTB board also hosts a set of ADC, DAC and digital potentiometer integrated circuits that communicate with the RDO board using the I2C protocol. These devices enable monitoring of the ladder temperature, ladder current consumption (analog and digital), MTB output voltages, as well as remote control of the individual over-current protection thresholds and the ladder power supply voltages.

\subsection{Read-out board and firmware}
\begin{figure}[h!]
\centering 
\includegraphics[width=.85\textwidth]{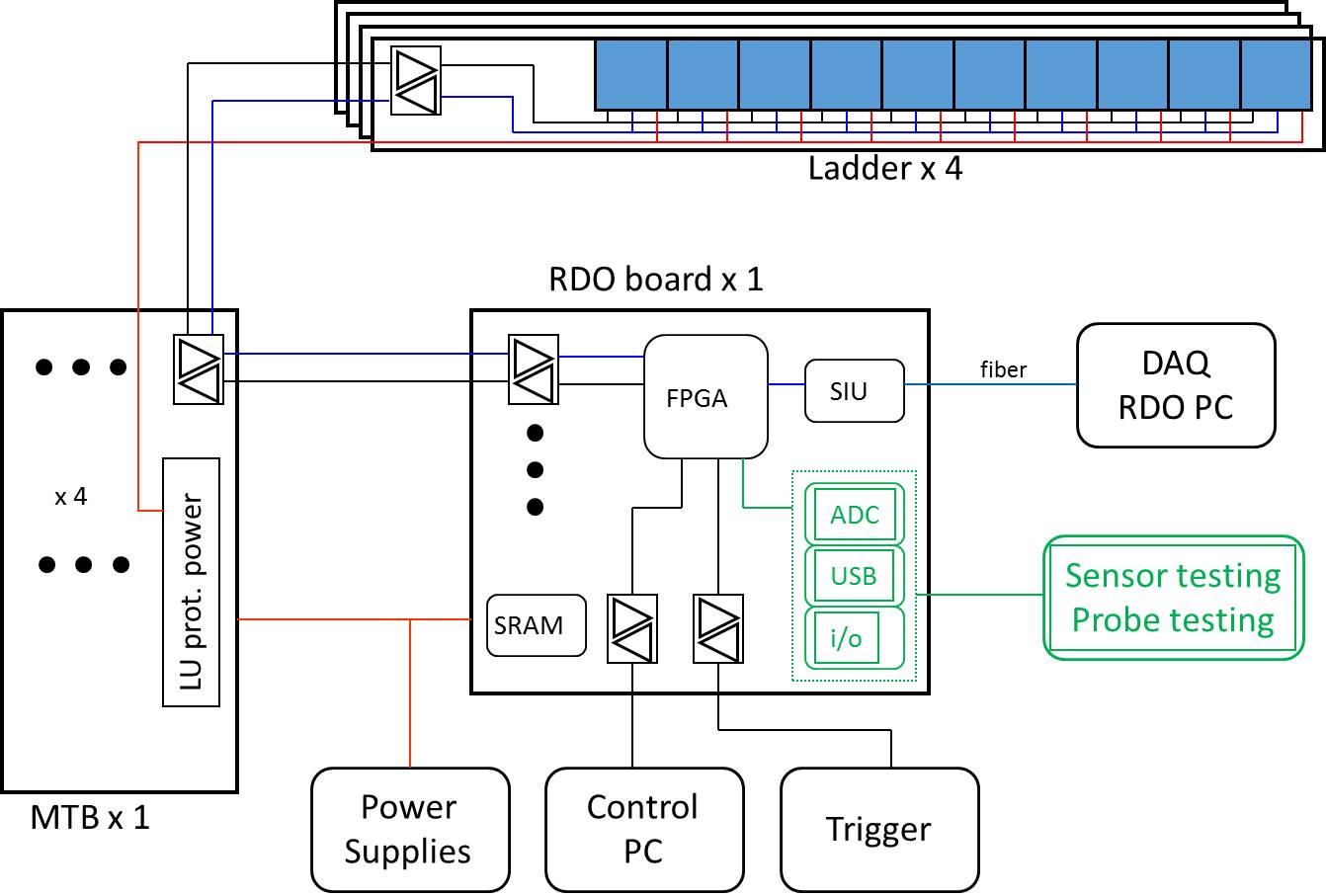}
\caption{\label{fig:RDOsys_blockdiagram}PXL readout architecture.}
\end{figure}

The FPGA based RDO board receives the data, performs trigger based hit selection, buffers, and formats the resulting data into event structures, and then sends it over 100 m optical fibers to one of ten fiber readout channels in the DAQ room. The interface to DAQ is accomplished over the bi-directional ``Detector Data Link'' (DDL) standard adopted by STAR from ALICE \cite{DDL}. The DDL standard consists of a daughter card called ``Source Interface Unit'' (SIU) on the RDO board side and a PCI-X based card with two bi-directional fiber interfaces called the ``Readout Receiver Card'' (RORC) on the PC side, connected via bi-directional fibers. The RORC cards are housed in standard server-type PCs in the DAQ room (up to 3 dual-channel RORCs per PC).
A block diagram showing the functional components of the RDO system and boards is shown in Figure \ref{fig:RDOsys_blockdiagram}.

The RDO board is also the interface to the trigger system and receives triggers from the STAR ``Trigger \& Clock Distribution'' (TCD) board \cite{trigger}. The RDO card can receive commands and data to configure and control the sensors either via the DDL link from the DAQ computer or over USB from a control PC. The USB link is also used to send monitoring data to the slow controls system of STAR. The configuration and control interface between the RDO board and the sensors uses the JTAG standard, via signals sent from the RDO board through the MTB to the ladders. The RDO boards are realized as 9U-size VME cards to be mechanically mounted in a VME crate. The VME P1 backplane is used to distribute the TCD signals to the 10 RDO boards via an interface board housed in the same VME crate. The TCD interface board converts the signals received from the TCD board for distribution to the RDO boards and also provides an \emph{OR} for the \emph{PXL-busy} feedback from the RDO boards to the trigger system. Each RDO board also contains daughter board connectors which provide the possibility to add additional circuitry for sensor testing and characterization, as well as additional prototyping. The FPGA chosen for the production version of the RDO board is from the LX240T sub-family of the Xilinx Virtex-6 FPGA (XC6LVX240T-FF1759). This FPGA provides powerful mixed mode clock managers for frequency synthesis, clock-phase shifting, and clock division, up to 15~Mbit of embedded dual port RAM, which can be used to facilitate the buffering of the data, and 720 User I/O pins. It incorporates Xilinx high-performance ``SelectIO'' technology, which allows individually controllable impedance active termination and signal delays for each I/O pin; each I/O pin has access to its own serializer and deserializer (``SerDes'') with programmable width and also supports double data rate (DDR) signaling. The FPGA can be configured from an on-board parallel flash or via JTAG either from a connector on the front of the board or from the USB interface. Connectivity to the USB bus is provided by a commercial interface board that converts the serial USB data stream into a FIFO-like parallel interface to the FPGA. The USB interface chosen on the RDO board is provided by the ``FT232H Mini Module'', which is a development module from the company ``Future Technology Devices International Limited'' (FTDI). This module utilizes the FT2232H USB Hi-Speed two-port bridge chip to handle all of the USB signaling and protocols compliant with USB 2.0 High Speed (480Mb/s), and has the capability of being configured in a variety of industry standard serial or parallel interfaces towards the RDO board, including asynchronous or synchronous parallel FIFO interfaces. Software drivers for the Windows and Linux operating systems are provided by the manufacturer. The interfaces to the four ladders of a sector are provided by high density connectors on the back of the RDO board.

There are several requirements on the firmware architecture that result from the RDO board design and the STAR infrastructure with respect to DAQ, trigger and slow controls. Each RDO board is connected to 4 ladders containing 10 sensors each, a total of 40 sensors per RDO board. The digital sensor data is passed through a zero-suppression block on the sensor, resulting in run-time encoded hit addresses. The zero-suppressed data are arranged into 16-bit words and transmitted out of the sensors serially (one bit per clock) over two LVDS lines in parallel. So each RDO board needs to receive a total of 80 LVDS lines from the sensor ladders, each running at 160Mb/s. After initial configuration of the sensors over JTAG is accomplished by the RDO board, the sensors are run continuously. Only after receipt of a Level-0 trigger from the TCD will the RDO boards generate an event, which must consist of data corresponding to a whole sensor frame relative to the trigger time. If multiple triggers occur within one sensor frame (~190~$\mu$s), each trigger must result in a separate event consisting of data for one frame, i.e. some data might be sent to DAQ multiple times as part of separate events. The RDO board needs to be able to handle average trigger rates up to 1~kHz. In order to not increase the DAQ dead time, the PXL RDO must match the burst capability of the TPC readout (the largest data generating source within STAR DAQ). The TPC can handle one event every 50~$\mu$s in short bursts of up to 8 events. For each trigger, the RDO board then formats the sensor data into an event structure with additional geographical information, and then sends these data over the fiber interface to DAQ. In order to adapt the sensor readout speed to the fiber speed, the RDO board needs to be able to buffer events locally. Simulations show that each sector (consisting of 10 sensors in the layer closest to the beam and 30 sensors in the next layer) on average has 10,000 hit pixels per integration frame. Only reading out hit pixels and efficient on-chip data encoding reduces the data rate from the ladders to the readout electronics significantly.
\begin{figure}[h!]
\centering 
\includegraphics[width=.85\textwidth]{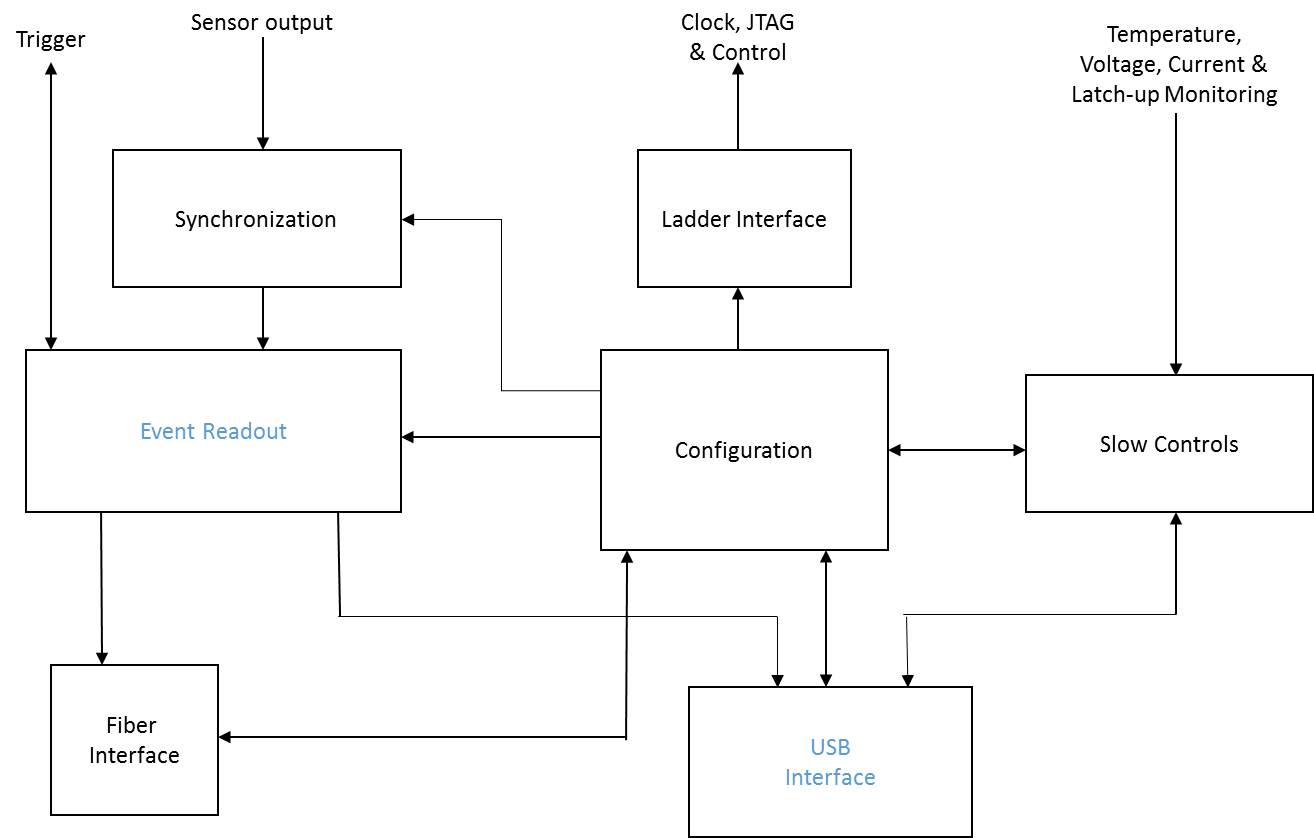}
\caption{\label{fig:firmware_blockdiagram}Firmware architecture block diagram.}
\end{figure}
Given these requirements, the firmware architecture is divided into several functional blocks shown schematically in Figure \ref{fig:firmware_blockdiagram}. A ``Configuration'' block deals with receiving RDO and ladder configuration data and commands either from the fiber or the USB interface. This block interfaces internally with the other firmware components to configure the requested RDO running conditions, and also interfaces with the sensors on the ladders via a JTAG control interface to set and read back JTAG configuration registers in the sensors that determine the operational parameters of the sensors. This firmware block also contains configuration memory so operational parameters can quickly be reloaded into the sensors without the need to send them from a PC again. A ``Slow Controls'' block reads temperatures and other operational parameters from the ladders and the MTB, monitors power latch-ups on the MTB and transmits these data to a slow controls PC via the USB interface. It also collects performance statistics from the rest of the RDO firmware to be recorded by the slow controls system.

\begin{figure}[h!]
\centering 
\includegraphics[width=.85\textwidth]{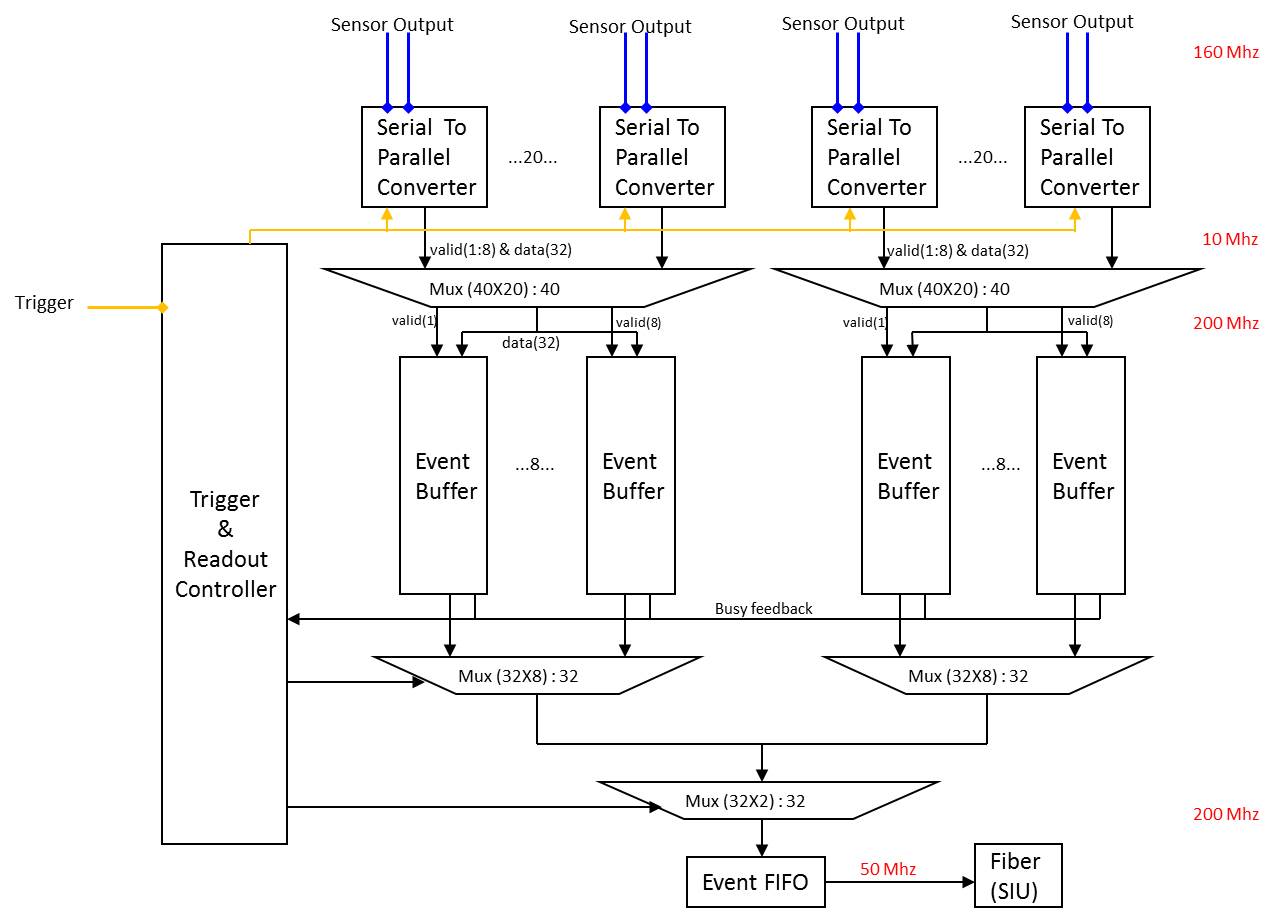}
\caption{\label{fig:EventReadoutModule}``Event readout'' module functional components.}
\end{figure}

The central and most complicated firmware block, ``Event readout'', deals with the event readout from the ladders in response to STAR triggers; a block diagram showing the functional components of this firmware block is shown in Fig. \ref{fig:EventReadoutModule}.

Data from the 40 sensors of a sector arrive in 80 differential serial lines at the FPGA, two per sensor. In the ``Synchronization'' firmware block (shown in Figure \ref{fig:firmware_blockdiagram}) the input signals are first individually aligned to the clock edge for proper sampling, using a feature of the Virtex-6 FPGA family that provides programmable delays for each individual I/O pin of the FPGA (``IODELAY''). After this initial signal alignment, the bits arriving on the two serial lines from each sensor need to be combined into 32-bit data words.
This is done in 40 parallel ``Serial-To-Parallel Converter'' (S2PC) firmware modules. A signal received from one of the sensors on each sector provides the frame synchronization of all of these modules relative to a common clock. Trigger information is received and processed in the ``Trigger \& Readout Controller'' (TRC) module of the firmware. When a valid Level-0 trigger is received by this module, it determines the next available event buffer to store the frame data associated with this trigger. In order to match the burst capability of the TPC, the firmware allows the data to be stored in one of 8 event buffers. In order to match the resource distribution within the FPGA, two sets of 8 event buffers are provided, where each set provides storage for 20 of the sensors. If several triggers are currently being processed, the same data coming from the S2PC modules will be stored in multiple event buffers, each buffer corresponding to one trigger. The TRC keeps track of each trigger currently being processed, and makes the corresponding event buffer unavailable for new triggers until the current event has been completely processed and passed to the next stage in the firmware. If all event buffers are currently processing events, the TRC sends a busy signal (combined together with busy signals from the other RDO boards with an OR operator) back to the trigger system, thus preventing additional event triggers until at least one of the event buffers is available again. The Level-0 trigger information as well as other trigger data received are stored locally in the TRC module and are appended to a complete event for further processing in the DAQ. After all of the data for one event has been stored in a set of event buffers, the TRC module combines the data from the two event buffers corresponding to one trigger and sends the data into an ``Event FIFO'', together with the currently available trigger data. A ``Fiber Interface'' firmware module reads the Event FIFO and sends the data to the SIU, which in turn uses its own firmware to send the data to DAQ. The FPGA chosen for the RDO board contains sufficient amounts of memory resources to provide all of the above described buffers and all necessary FIFOs and elasticity buffers internal to the FPGA, thus greatly reducing the firmware complexity.

All of the above described firmware modules are necessary during normal data taking mode in STAR. The same hardware 
has also been used during ladder testing and characterization in the lab
. These test modes often don’t require the high data rates necessary in the STAR experimental setup, so that data readout via the USB interface is sufficient. This readout mode can also be used during the commissioning phase or other debugging operations, when DAQ readout over the fiber might not be available. 

\section{Mechanics}\label{sec:mechanics}

\subsection{Mechanical structure}
The PXL detector is the innermost and has the highest single point resolution of the detectors of the HFT upgrade. Correspondingly, it needs to be supported in a stable and repeatable manner (under multiple insertions) to provide the full upgrade with the desired DCA pointing resolution \cite{howard}. The primary design goals for the PXL mechanics include:
\begin{itemize}
\item The pixel positions shall maintain their measured positions to a 20~$\mu$m tolerance envelope under all operating conditions.
\item The PXL detector shall be provided with a rapid insertion and removal mechanism that allows the full detector to be removed and a duplicate spare detector to be inserted and operational within 24 hours maintaining the metrological map of the spare detector.
\item The PXL detector shall be designed in such a manner that the pixel positions can be accurately measured on a complete half detector including in the sensor overlap regions.
\item The radiation length of the supports shall be less than 0.15\% in the radial direction for perpendicular incidence.
\end{itemize}
The design specification for the PXL detector support mechanics and the PXL sectors were generated such that the position of each pixel in the detector should hold its measured position to a 20~$\mu$m tolerance envelope under all of the forces that arise from detector assembly and operation. This requires that the forces arising from the air cooling flow (10~m/s), temperature variations, thermal increase during operation, humidity absorption, and vibration need to be very well controlled and distortions mitigated. The insertion mechanism must be both easily operable and provide for a well-controlled locking mechanism that allows for the detector halves to be registered into a known position reproducibly. The overall design of the individual sectors and half detectors must be designed in a manner that accounts for the limitations of metrological mapping on a Coordinate Measuring Machine (CMM). These requirements and the accompanying designs will be described in order. 

\subsection{Sector design}
PXL sectors are designed to be rigid structures that maintain the PXL sensor ladders in well-defined positions. This is accomplished by designing the main structural component, the sector tube, to be a rigid low mass shell that maintains pixel positions under cooling airflow and gravity sag. The sector tube design is composed of an ALICE style trapezoidal tube fabricated with 7 layers of Toray M55J high-modulus unidirectional fiber prepreg with cyanate ester resin in a 0,-60,+60,0,+60,-60,0 deg layup to give a final thickness of 120~$\mu$m. The large sectional cross section between the inner and outer ladder mounting surfaces provides for excellent stiffness in the radial direction. 

\begin{figure}[h!]
\centering 
\includegraphics[width=.85\textwidth]{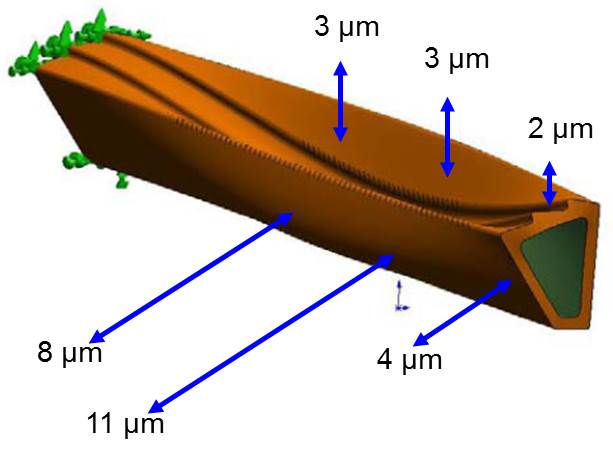}
\caption{\label{fig:PXL_FEA_deformation} The lowest order vibrational mode was calculated to be 259 Hz with FEA and measured to be 230 Hz on the final production sectors. The deformations are shown as the distorted image of the of the sector tube. The measure static deformation numbers shown were obtained using a capacitive probe while exposing a sector tube to 9m/s air flow. This is within the required tolerance window.}
\end{figure}

Ladders are mounted onto the three top flat rectangular regions and onto the bottom surface. This provides the two tracking layers required. Finite element analysis (FEA) was done for the sector tubes to show expected vibrational modes that could result from air flow. Actual measurements of vibration and static deformations were accomplished using a capacitive probe on a sector tube exposed to the 9~m/s airflow. The measured static deformation is shown in Figure \ref{fig:PXL_FEA_deformation}. The sector deformation due to gravitational sag for any sector position has a maximum of 4~$\mu$m on the sector sidewall and less than 1 micron in the ladder positions. 

\subsection{Insertion mechanism design}
The PXL detector when in the home position in the STAR detector for data taking, is located with the STAR beam interaction diamond located in the geometric center of the PXL ladders. The PXL detector is precisely placed in this location through the use of a set of carbon fiber rails mounted inside the carbon fiber cylindrical shell surrounding the detector. When the detector has moved into the proper position along the beam axis, the PXL detector is locked into a defined and repeatable position using a set of kinematic mounts mounted in the PXL support carbon fiber cylindrical shell. The PXL detector mechanics was designed with the goal of allowing for rapid insertion and detector replacement within 12 hours. This was achieved using an insertion rail design in which each of the detector halves could be independently inserted and removed from the interaction region of the STAR detector using the method described. In addition, the full infrastructure needed for operating the PXL detector in the home position (MTBs, power distribution, etc.) were included as part of the pre-cabled and tested assembly that inserts along the rails with each PXL detector half.

\begin{figure}[h!]
\centering 
\includegraphics[width=.75\textwidth]{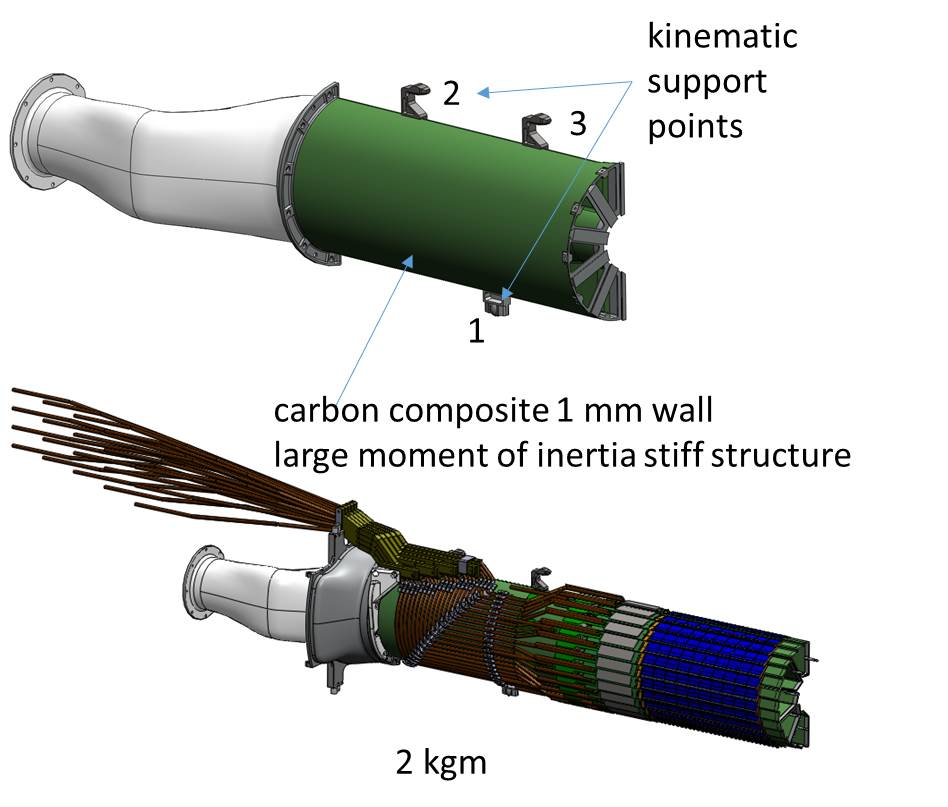}
\caption{\label{fig:PXL_half_support} The PXL detector half support structure. The upper drawing shows the kinematic mounts located on the support structure. The lower diagram shows the support structure with the sectors mounted and cabling services routed out. The mass of the fully cabled detector half with sectors and cabling is $\sim$2~kg.}
\end{figure}

The primary support for the detector ladders is the detector half shell (see Figure \ref{fig:PXL_half_support}). This piece holds the sectors into the proper positions using a set of precision machined mounts and then locates the detector half assembly precisely inside the interaction region with the attached kinematic mounts. The sector tubes terminate in precision machined set of wedge shaped aluminum rails that mate with and lock into  to a complementary set of rails machined into the face plates mounted to the half shell cylinder. The positioning accuracy and stability and reproducibility  provided by this arrangement is very high (order of a few $\mu$m).

\begin{figure}[h!]
\centering 
\includegraphics[width=.95\textwidth]{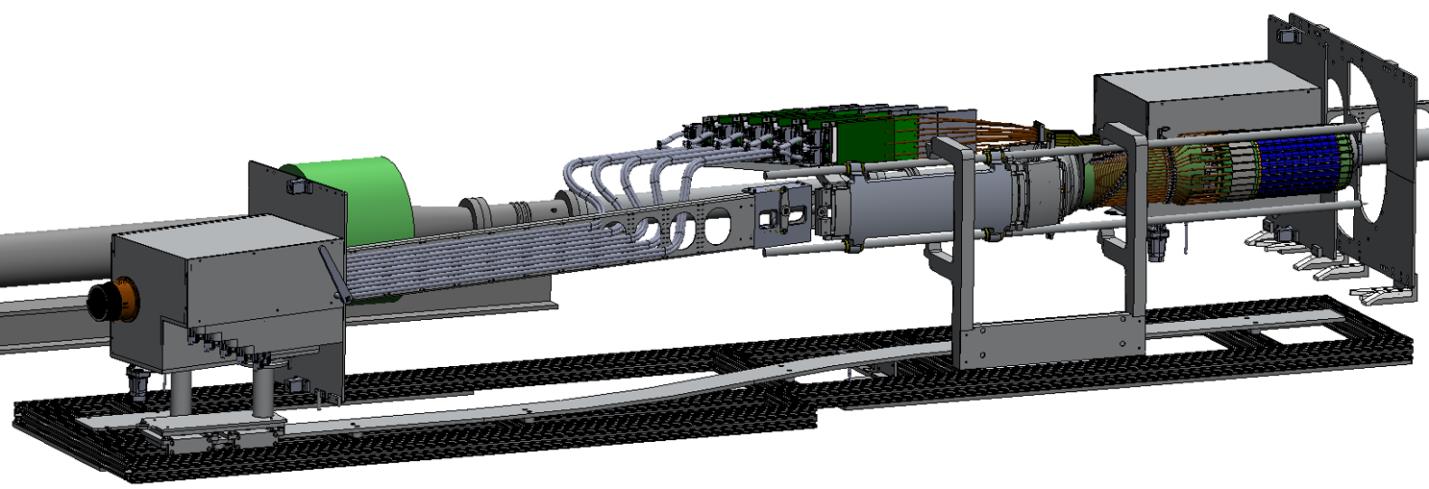}
\caption{\label{fig:PXL_transport_carriage}PXL detector half on transport carriage ready to be inserted into the STAR detector. Note that the services (mass termination boards, power distribution, etc. are integrated into the fully tested package that is inserted into the STAR inner field cage.}
\end{figure}

These primary sector support pieces were mounted to a set of hinged rail guides that then were slid onto the rails of a PXL transport carriage which could be placed as a unit at the entrance of the STAR TPC field cage (see Figure \ref{fig:PXL_transport_carriage}). The rails on the PXL transport carriage were aligned to the rails mounted to the carbon fiber polycaprolactone (PCL) support cylinder and the detector half was then slid into place. In order to support and align the PXL transport carriages, a 6 degree of freedom jack screw table was designed and fabricated. This arrangement allowed for a PXL detector half stored on the rails of a transport carriage to be put into position on the external alignment system and then properly aligned and inserted into the rails in the PXL support shell. 

\begin{figure}[h!]
\centering 
\includegraphics[width=.85\textwidth]{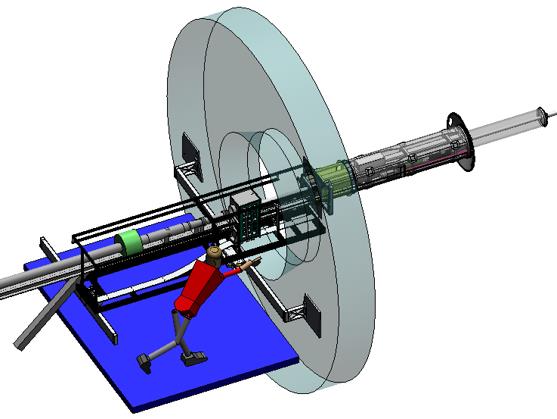}
\caption{\label{fig:PXL_detector_pushed}The PXL detector halves are manually pushed into the home position along the rails mounted in the PXL insertion cylinder.}
\end{figure}

Once aligned, the PXL detector halves are manually pushed along the rails into the home position and locked in place with the kinematic mounts (Figure \ref{fig:PXL_detector_pushed}).

\begin{figure}[h!]
\centering 
\includegraphics[width=.85\textwidth]{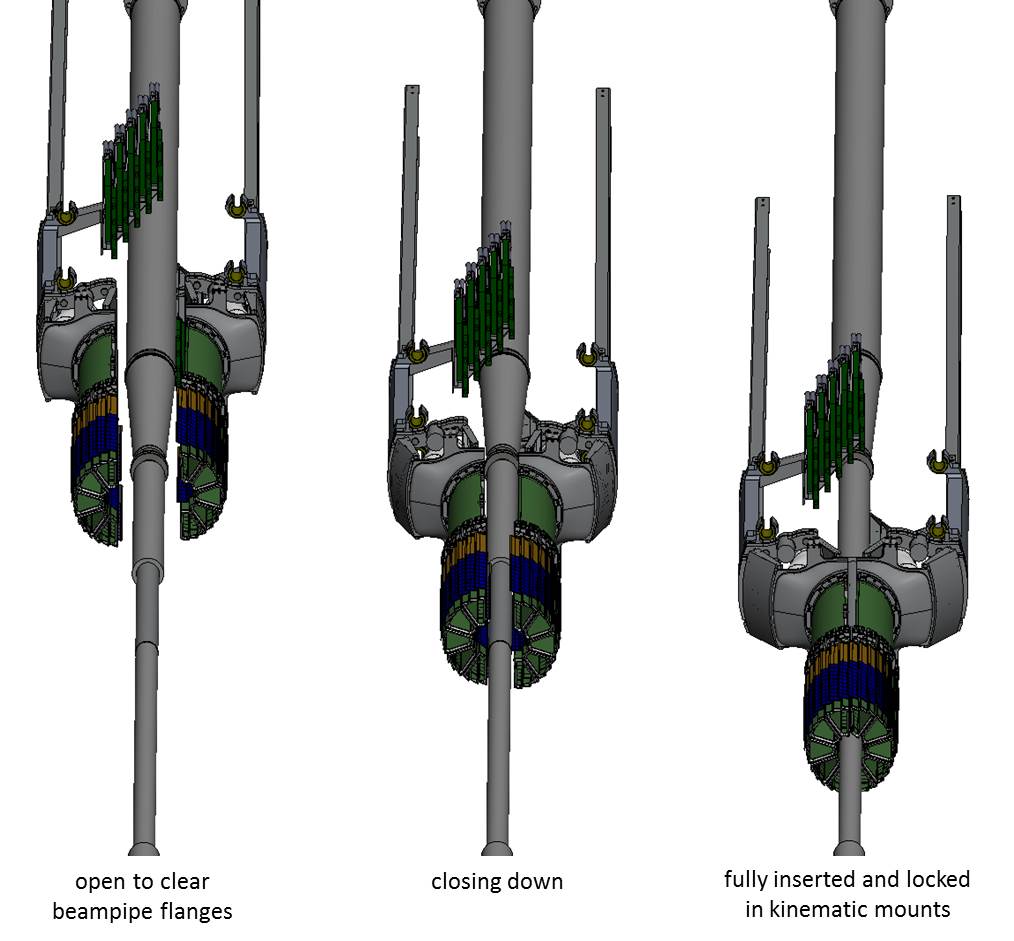}
\caption{\label{fig:PXL_halves_guided} The PXL detector halves are guided into a closed position using guide rails that bring the PXL detector halves into a closed hermetic position where they are locked into kinematic mounts.}
\end{figure}

The insertion of the PXL detector halves into the home position is guided along a path that closes the detector half around the beam pipe using a set of guide rails that define the insertion path as it is inserted. This gives the detector radial hermetic coverage around the beam pipe and interaction region. This is illustrated in Figure \ref{fig:PXL_halves_guided}.

The PXL mechanical systems worked very well and during running conditions a full PXL detector swap into and out of the STAR detector was achieved in a single day.

Repeated insertions into the kinematic mounts led to a detector position that was stable to within 6~$\mu$m FWHM with the PXL detector position within the range of expected positions based on the manufacturing tolerances of the support pieces. The positional relationship between the PXL, IST and SSD were set by the mechanical properties and fabrication tolerances of the carbon fiber cylinders and mounting points that were used by all three detectors. While the absolute positioning of the relative detector positions was maintained to a few hundred micron precision, the most important characteristic of the mechanics was the stability of the relative positions of the detectors, which was maintained to a 50~$\mu$m envelope. Note that the pointing resolution of the outer detectors is of the order of a few hundred microns.
The clearance between adjacent sectors in each of the the detector halves in the closed position is the same as clearance between the sectors in a detector half. The distance of closest approach of sector assembly edges occurs between the outer edge of the driver board area of the ladder and the components mounted to the adjacent ladder driver board with a clearance of approximately 0.5~mm.

\subsection{Cooling system}
\begin{figure}[h!]
\centering 
\includegraphics[width=.95\textwidth]{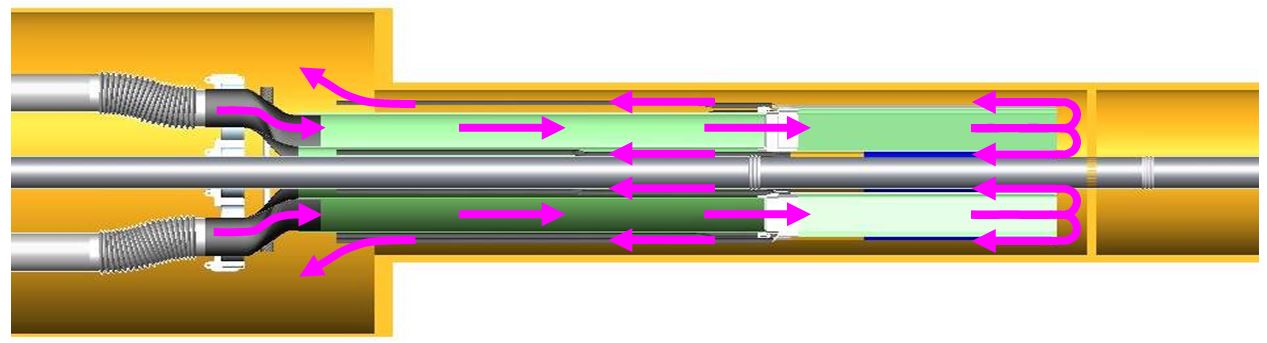}
\caption{\label{fig:PXL_airflow_cooling} Graphical representation of air flow in the PXL cooling system.}
\end{figure}

Air cooling has been chosen for the PXL detector in order to minimize multiple coulomb scattering.  A number of studies were carried out to optimize and validate the air cooling design, in which the incoming airflow is directed through the inside of the sector tubes, constrained by a barrier in the PXL support cylinder, and then returns and exits the detector over the outer surface of the ladders and along the inner barrel surface next to the beam pipe. This is shown graphically in Figure \ref{fig:PXL_airflow_cooling}.
 
The pixel chips dissipate 150~mW/cm$^2$ or a total of approximately 270 watts and additional 80~W are required for the ladder drivers. In addition to the ladder total of 350~W a fraction of this value is required for voltage regulators and electronics that are on the MTBs and reside in the air cooled volume in the exit path of the airflow. 
 
Initial computational fluid dynamic simulations indicated that the temperature rise of the sensors could be limited to approximately 10$^{\circ}$C above ambient temperature with a cooling airflow of 9~m/s. This result was later confirmed in cooling test studies using a full-size detector mock-up as described below. 
 
Since MAPS-based devices can be operated at room temperatures and up to $\sim$40$^{\circ}$C without any significant noise degradation, the PXL detector is cooled simply with forced air flow and without any need for thermal isolation and condensation control.
The air chiller system\footnote{Water-Cooled Environmental Control Unit (ECU), Air Innovations Model 99C0134-00, https://airinnovations.com/} provides the cooling air circulating through the PXL detector with a temperature of around 23$^{\circ}$C as measured at the blower outlet. 
The chiller and the PXL support structure are connected with 6~inch (15.24~cm diameter) flexible ducts. The air exiting the PXL detector volume is released into the STAR hall. 

\begin{figure}[h!]
\centering 
\includegraphics[width=.85\textwidth]{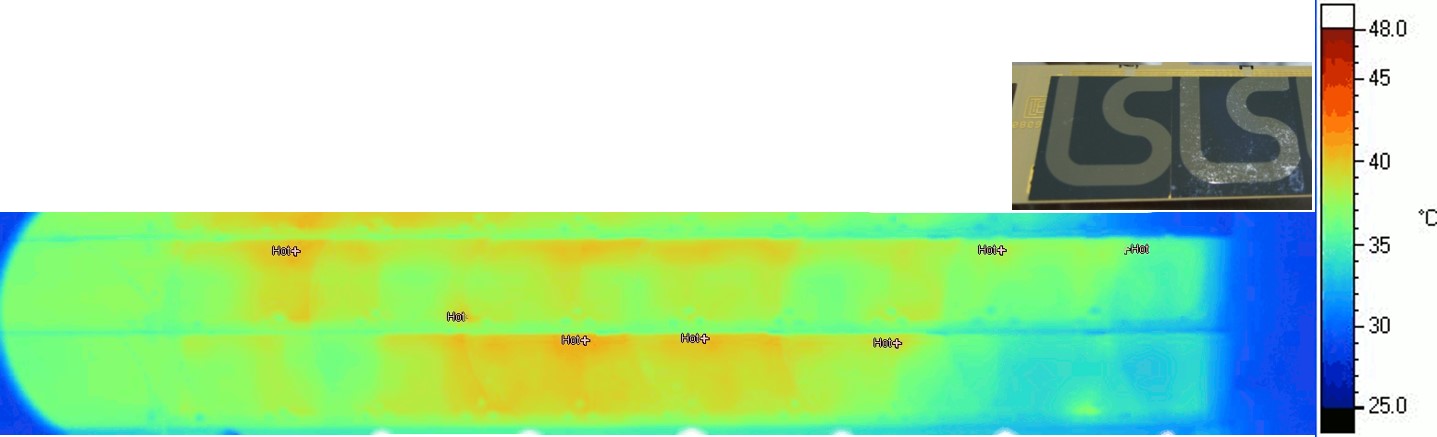}
\caption{\label{fig:PXL_composite_thermal}Composite thermal image of the temperature profile over the outer ladders in a sector for 10.1~m/s of cooling airflow. This test was performed using platinum heaters vapor deposited onto 50 $\mu$m silicon blanks shown in the upper right photograph.The full ladder length of 10 sensors (20~cm) is shown. The thermal dissipation is 150~mW/cm$^2$. }
\end{figure}

Initial testing was done with sensor mock-ups consisting of vapor deposited platinum heater elements on 50~$\mu$m silicon blanks (see Figure \ref{fig:PXL_composite_thermal}). This testing showed that with the expected power dissipation of 150~mW/cm$^2$ in the sector, the temperature rise in the middle of the ladder (hottest point) for a measured 10.1~m/s of airflow resulted in a 10 degree rise above ambient. There were no significant differences in temperature or profile measured for the inner ladder versus the outer three.

\begin{figure}[h!]
\centering 
\includegraphics[width=.75\textwidth]{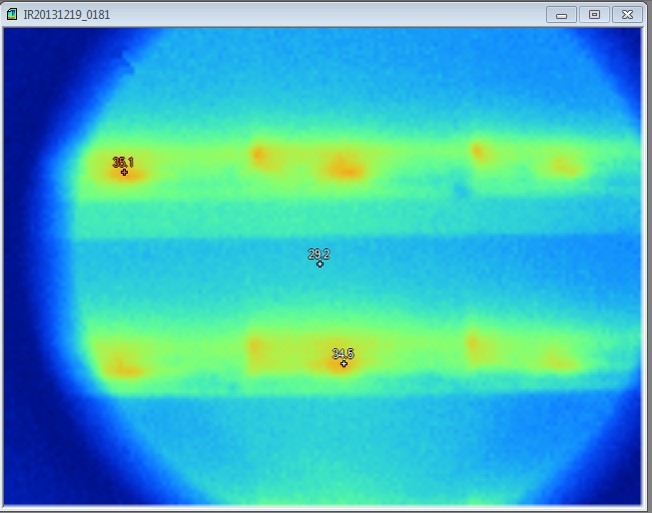}
\caption{\label{fig:PXL_sector_thermal} Thermal image of two production sectors in a truncated production PXL support tube showing the maximum temperature of 36.1~$^{\circ}$C on the sensors corresponding to a 12$^{\circ}$C maximum sensor temperature rise at the nominal sensor dissipation of 170 mW/cm$^2$. Within a sensor, the digital periphery shows a temperature about 6$^{\circ}$C higher than the rest. Due to limitations in the infrared viewing port size, 8 sensors are shown in the picture (16 cm of length).} 
\end{figure}

This testing was repeated with the production detector as a final check of the cooling system developed. A thermal image of production sensors under standard operating conditions is shown in Figure \ref{fig:PXL_sector_thermal}. The cooling airflow is regulated to be 23.1$\pm$1$^{\circ}$C. The maximum temperature on a sensor is 36.1~$^{\circ}$C giving a maximum temperature rise of 13$^{\circ}$C. In these cooling conditions, the temperature variation within a sensor is about 6$^{\circ}$C, with the maximum temperature measured in the digital periphery.

\subsection{Vibration and displacement testing}
The mechanical system testing for PXL was extensive as it was required that the pixel positions be known and stable during operating conditions including 10 m/s cooling airflow.  Most of the initial design of the PXL mechanical components included a full set of FEA analysis but the most critical parts needed prototyping and testing to be sure that they would perform as expected. We describe the validation testing used on prototypes and production design parts to verify that the critical components were performing to the specification \cite{howard}.

\begin{figure}[htbp]
\centering 
\includegraphics[width=.445\textwidth]{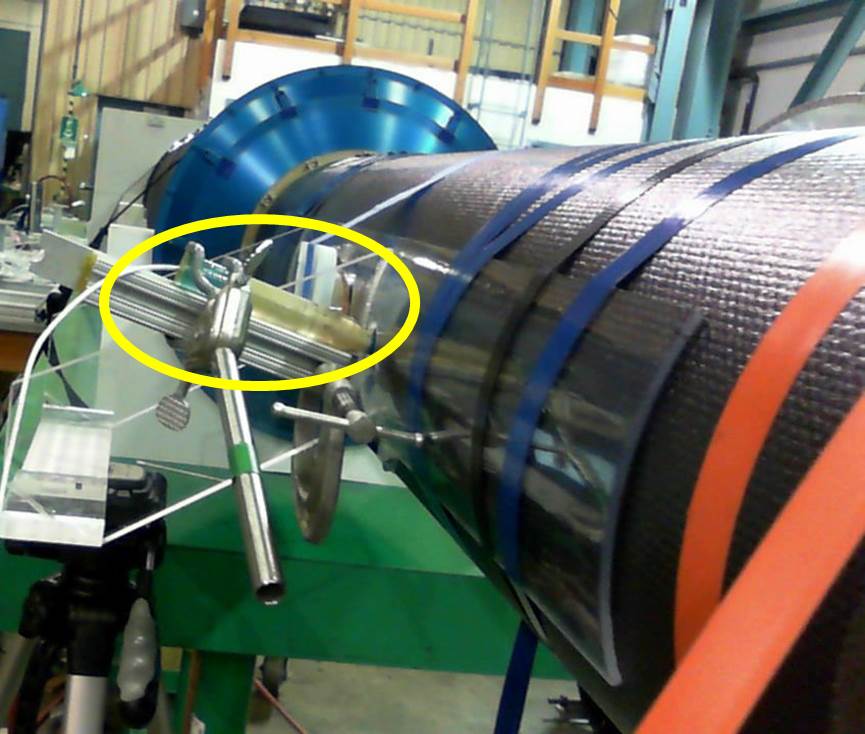}
\qquad
\includegraphics[width=.43\textwidth]{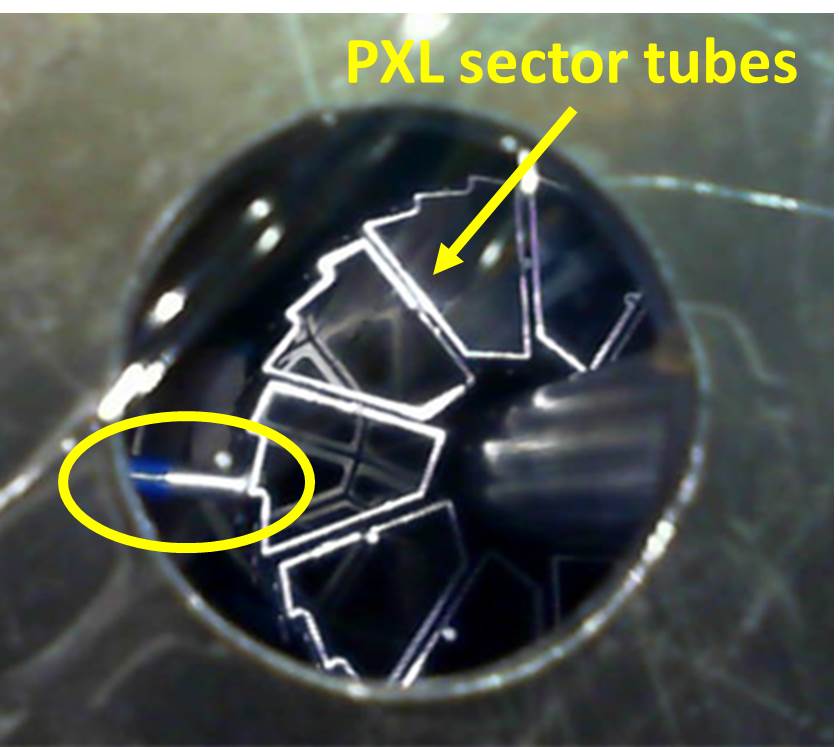}
\caption{\label{fig:PXL_vibration_testing}Left: Picture showing the setup of the final testing of the vibration and displacement in the operating conditions using production sensors and mechanics with the PXL detector locked into the kinematic mounts and inserted into the PXL support tube to direct the cooling airflow. The capacitive probe is circled in yellow in both pictures. Right: the non-touch capacitive probe used to measure displacement and vibration is positioned close to the surface of the ladder near the 9 o'clock position. The end of the sector tubes in the sector half are visible}
\end{figure}

In an airflow environment of 10 m/s, vibration and displacement of the sectors is a concern that needs to be addressed. An initial set of displacement and vibration measurements were made at the same time as the initial thermal testing with the platinum heater sensor mock-ups, but the sector mounting supports were fabricated from rapid prototype plastic and the measurements, while showing that the vibrations were at the limit of the acceptable range, were not considered reliable. A more comprehensive set of tests was done with the final kinematic mounts and production aluminum sector mounting supports. The testing setup is shown in Figure \ref{fig:PXL_vibration_testing}.

The test results are shown in Table \ref{tab:Mechanicstest}.

\begin{table}[h!]
\begin{center}
  \begin{tabular}{|  l  |  l | }

\hline
Sector fundamental resonance frequency:	& 230~Hz (FEA simulation gave 260~Hz) \\ 
Sector vibration at full flow: &   5~$\mu$m RMS  \\ 
Sector displacement at full flow (inward): &  25-30~$\mu$m  \\ 
\hline
    \end{tabular}
    \caption{Airflow induced vibration and displacement test results.}
    \label{tab:Mechanicstest}
    \end{center}
\end{table}

\section{Detector production}\label{sec:production}
The PXL detector production began in Summer 2013
. The production work-flow and timeline are summarized in the next sections.

\subsection{Assembly}
\begin{figure}[!htb]
     \centering
     \begin{minipage}[b]{0.43\textwidth}
       \begin{subfigure}[b]{\linewidth}
     	\includegraphics[width=1\textwidth]{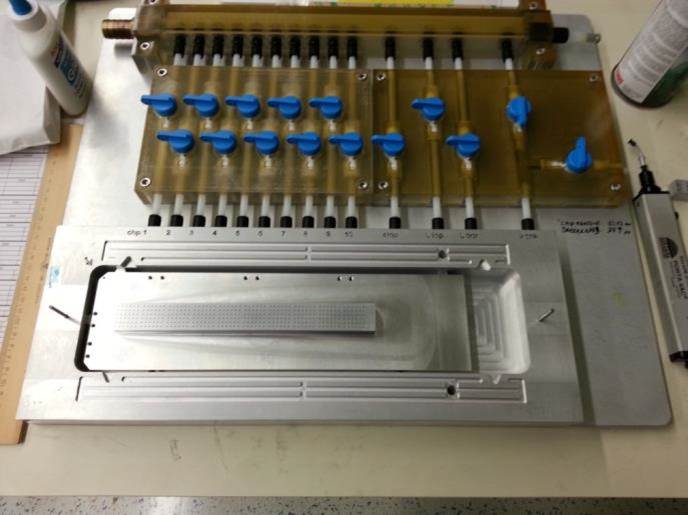}
       \caption{}
 		\label{fig:ladassa}
       \end{subfigure}\\[\baselineskip]
       \renewcommand{\thesubfigure}{c}
       \begin{subfigure}[b]{\linewidth}
         \includegraphics[width=1\textwidth]{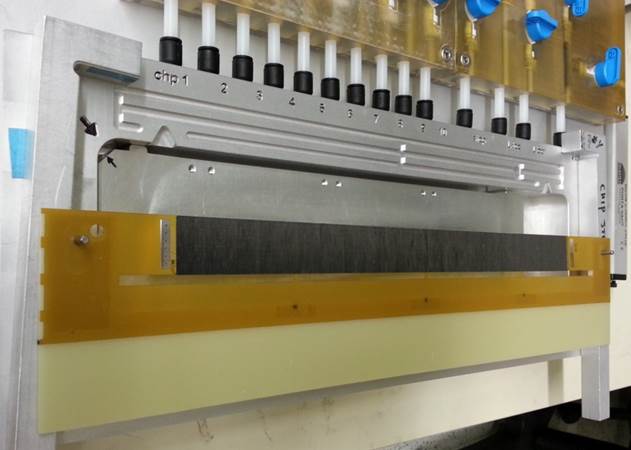}
         \caption{}\label{fig:ladassb}
       \end{subfigure}
     \end{minipage}
     \hfill
     \renewcommand{\thesubfigure}{b}
     \begin{subfigure}[b]{0.45\textwidth}
       \includegraphics[width=1\textwidth]{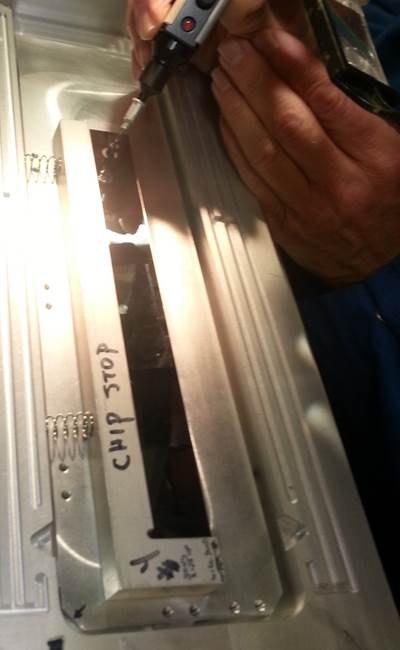}
       \caption{}\label{fig:ladassc}
     \end{subfigure}
     \caption{Ladder assembly. \emph{a)} Precision vacuum chuck fixture. \emph{b)} Sensors are positioned with butted edges on the fixtures. \emph{c)}The structure composed by cable and carbon fiber stiffener is aligned with the sensors through a series of pins and holes on the fixture.}\label{fig:ladder_assembly}
   \end{figure}

The PXL detector production consists of three subsequent phases: ladder assembly, sector assembly and detector-half assembly \cite{iworid}. All the components are validated prior to assembly and the functionalities of the manufactured devices are verified after each production step. A description of the sensor Quality Assurance (QA) probe test is given in Section \ref{sec:sensorQA}.
The finished sensor dimensions are inscribed as trenches on the surface of the wafers using a Deep-Reactive-Ion-Etching (DRIE) technique.  The sensors are then thinned by back grinding and polishing into the trench depth releasing the individual sensors. With DRIE, the trench is part of the lithography process of the wafer and has the associated precision, which is of the same order as the feature size of the process 0.35~$\mu$m. In the saw based dicing (which was also tested) the accuracy we achieved was approximately 5~$\mu$m with the chipping of the silicon in the dicing lanes that required maintaining a 25~$\mu$m boundary outside the sealing ring. The uniformity of the DRIE edge showed no observable chipping or flaws. The sensors are then fully probe tested and characterized and are manually positioned with butted edges to flex PCBs using precision vacuum chucks, as shown in Figure \ref{fig:ladder_assembly}. 
The sensors and the front-end electronics are attached to a flex cable with a very low elastic modulus acrylic adhesive\footnote{3M$^{\texttrademark}$ Adhesive Transfer Tape 467MP, \href{https://www.3m.com/3M/en_US/company-us/all-3m-products/~/3M-Adhesive-Transfer-Tape-467MP?N=5002385+3293242532&rt=rud}{https://www.3m.com/}}. This decouples elements with differing coefficients of thermal expansion for the purpose to prevent thermostat mode mechanical displacements that would exceed the required pixel position stability. The low modulus adhesive’s relieving of differential expansion stresses also prevents potential breakage of the thinned silicon detector chips. They are then electrically connected via standard wire bonding to the flex cable and the wires are encapsulated with UV-curable encapsulant\footnote{Dymax Multi-Cure$^{\tiny{\textregistered}}$ 9001-E-v3.1, \href{https://www.dymax.com/index.php/adhesives/products/9001-e-v31}{https://www.dymax.com/index.php/adhesives/products/9001-e-v31}} for protection. The structure is stiffened by a $\sim$125~$\mu$m thick carbon fiber backer glued by means of another layer of adhesive tape at the bottom of the flex cable. 

\begin{figure}
\centering
   \begin{subfigure}[b]{0.85\textwidth}
   \includegraphics[width=1\linewidth]{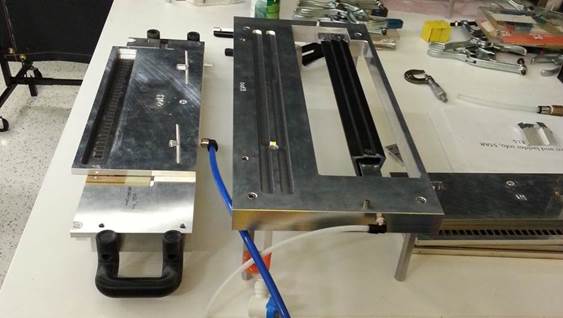}
   \caption{}
   \label{fig:Ng1} 
\end{subfigure}
\begin{subfigure}[b]{0.85\textwidth}
   \includegraphics[width=1\linewidth]{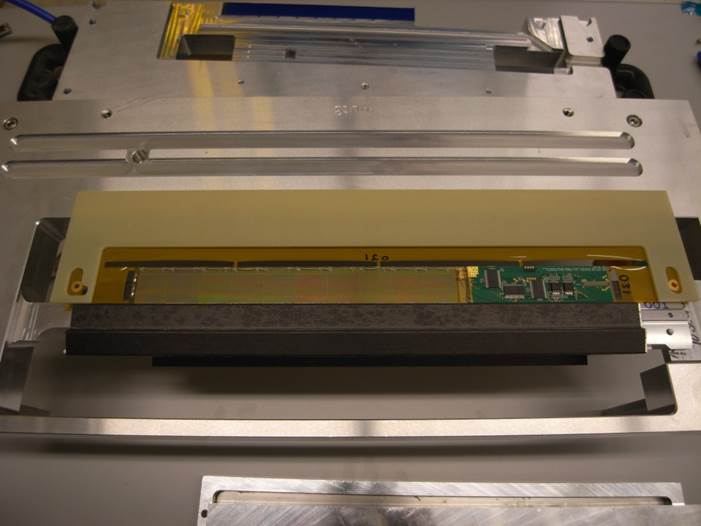}
   \caption{}
   \label{fig:Ng2}
\end{subfigure}
\caption{\label{fig:sector_assembly}Sector assembly: \emph{a)} A carbon fiber sector tube locked in position on the sector assembly fixture (\emph{right}). The glue will be applied on the ladder back side, positioned on the top chuck (\emph{left}). Afterwards, the ladder is flipped on the sector tube and glued on the exposed flat rectangular region. \emph{b)} After curing, the FR-4 handling frame is removed and the sector can be moved to the next sector assembly station.}
\end{figure}

\begin{figure}[h!]
\centering 
\includegraphics[width=.97\textwidth]{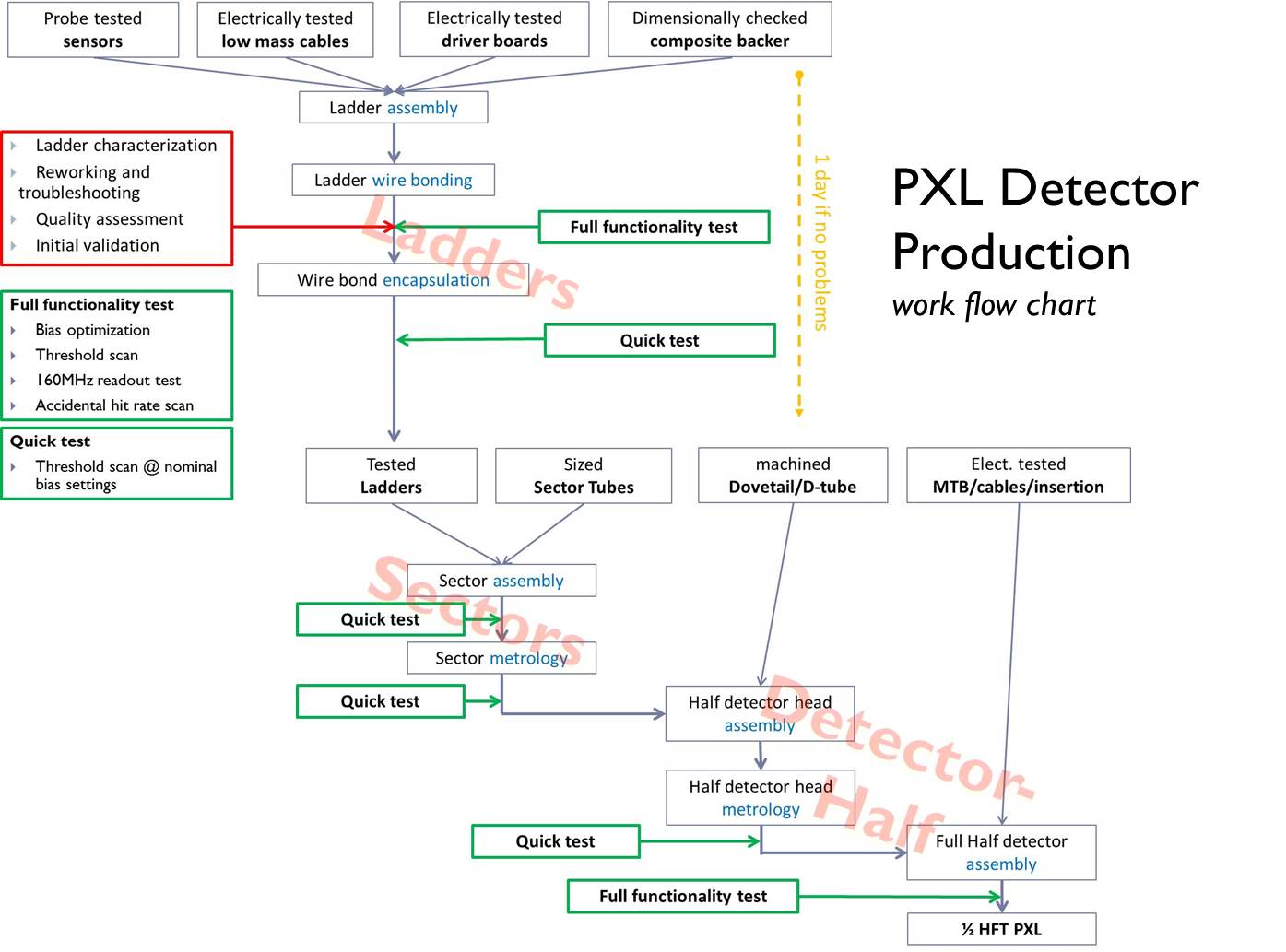}
\caption{\label{fig:workflow}The PXL production procedure. The work flow diagram summarizes the ladder assembly, sector assembly and detector-half assembly phases, and the quality assurance testing implementation.}
\end{figure}

At this point, precision vacuum chucks are used to position the ladders on the sector tubes and to glue them using silicone adhesive\footnote{APTEK 2724-A/B, produced by Aptek Laboratories Inc. (\href{http://www.apteklabs.com/}{http://www.apteklabs.com/}), is a two component, white, soft, thixotropic, electrically insulative, silicone adhesive displaying excellent flow temperature flexibility}. The ladder and sector assembly fixtures rely on a series of pins and holes for the alignment of the different components. Weights are taken at all assembly steps to track the material contributions and as QA. The typical weight for a complete ladder (sector) is $\sim$14~g ($\sim$100~g). Selected sector assembly phases are shown in Figure \ref{fig:sector_assembly}.
Fully assembled sectors are surveyed in a CMM, as described in Section \ref{sec:metrology}. The completed sectors are then inserted and locked into the sector mount joints to form detector-halves. Each half is surveyed to form completely mapped stable units with the pixel positions known to an accuracy of approximately 10~$\mu$m. Finally, the detector halves are attached to the insertion mechanics completing the detector assembly in preparation for installation into the STAR detector.
The assembly procedure is summarized in the work flow diagram in Figure \ref{fig:workflow}.

\subsection{Production timeline}
The detector production started in Summer 2013. The 40 top quality ladders, out of the 77 produced in 2013, were used to build the first PXL detector copy, delivered to BNL in January 2014. The primary detector counted less than two thousand unresponsive pixels out of $\sim$356 million. Due to delays in fabrication, only two aluminum conductor flex cables were featured in the inner ladders. All the other ladders composing the detector operated in the 2014 Run used copper conductor flex cables. The spare detector copy was assembled in Spring 2014, featuring aluminum conductor cables on all the inner ladders. During Fall 2014, the primary detector was refurbished after the damage induced by latch-up events in the 2014 Run operations, and to equip all the sectors with aluminum conductor on the Inner Layer. A total of 146 PXL ladders were produced. A power supply accident damaged one of the two detector copies during the 2015 Run installation. The other copy was installed and operated in the 2015 Run. The ladders damaged in the accident and the ones affected by latch-up induced damage in the 2015 Run were replaced in Summer 2015. In summary, a total of 127 ladders were installed on sectors during the various phases of construction/refurbishment of the PXL detector. The overall ladder assembly yield resulted to be $\sim$90\%. The production process required approximately 10 functionality tests for each ladder. This QA process is described in the next section.

\subsection{Ladder quality assurance}\label{sec:ladderQA}
The ladders are tested and characterized at each stage of the production. After wire-bonding the sensor pads to the flex conductor cable traces, the sensor functionalities can be characterized for the first time in the ladder assembly. A test setup based on the PXL prototype powering and readout system is used to control and configure the ladder, and readout the data. A Windows Batch Script based software controlled through a LabView GUI allows to power and configure up to 4 ladders (equivalent to a full PXL sector) at the same time and test them sequentially. An extensive characterization of the functional and performance parameters listed in Section \ref{sec:sensorQA} is performed immediately after the wire-bonding stage. Possible problems originated by the wire bonding can still be fixed before wire encapsulation. The wire bonding required to be partially repeated in about the 10\% of the cases to improve the quality of the ladders. A shorter test of a subset of parameters, consisting in a threshold scan at the nominal reference voltage, is repeated after the wire encapsulation phase and at each sector assembly step subset. The ladder is fully characterized again after completing the sector assembly and metrology survey. Once the detector-half is completed and surveyed, the PXL production powering and readout system is used to operate the sector units and to scan the sensor thresholds in full readout speed mode. For each ladder, the sensor test results have been automatically stored into a MySQL database and used for the detector configuration after the installation in STAR. 

\begin{figure}[htbp]
\centering 
\includegraphics[width=.46\textwidth, trim=1.0cm 0 1.2cm 0]{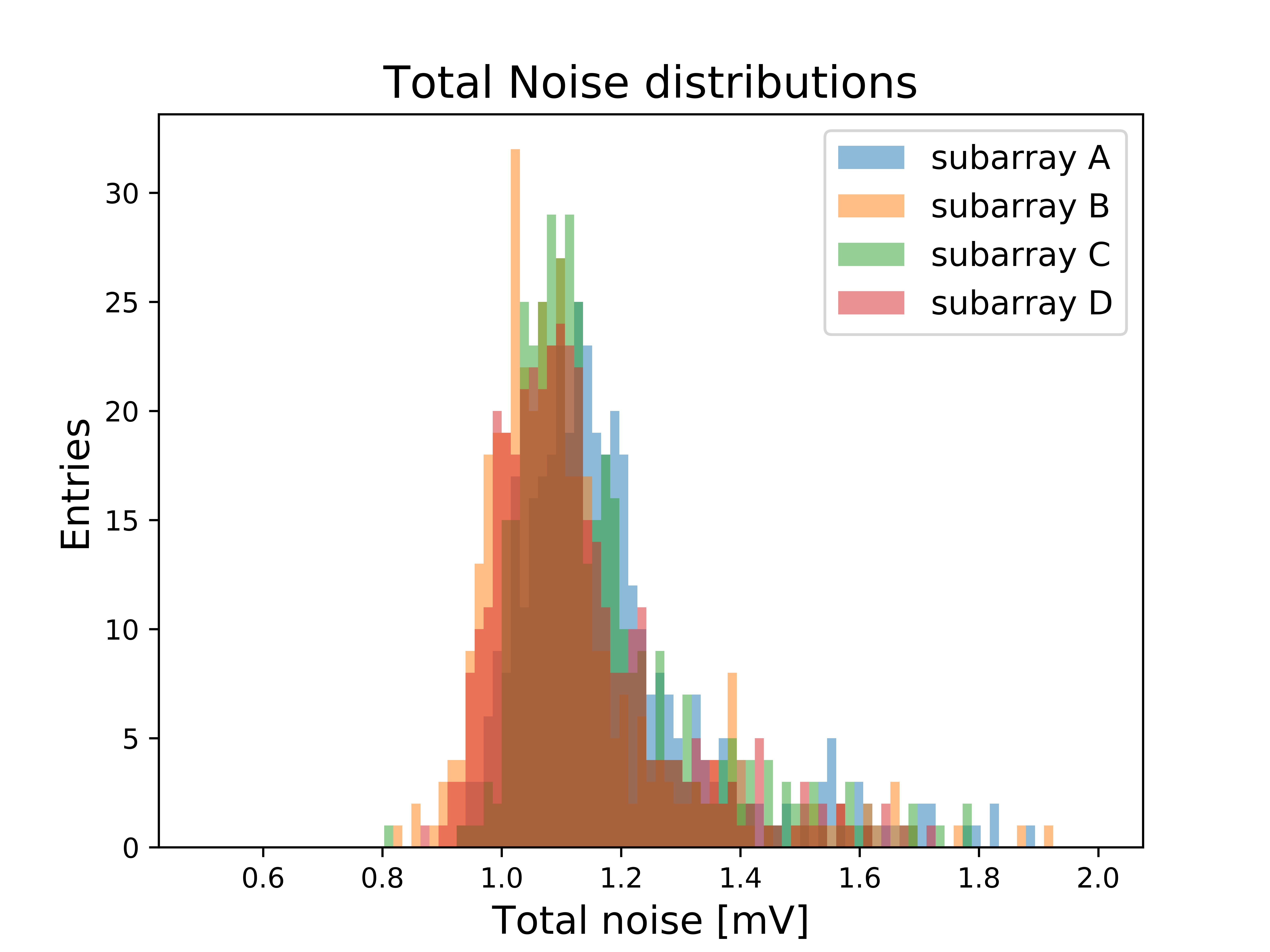}
\qquad
\includegraphics[width=.46\textwidth, trim=1.0cm 0 1.2cm 0]{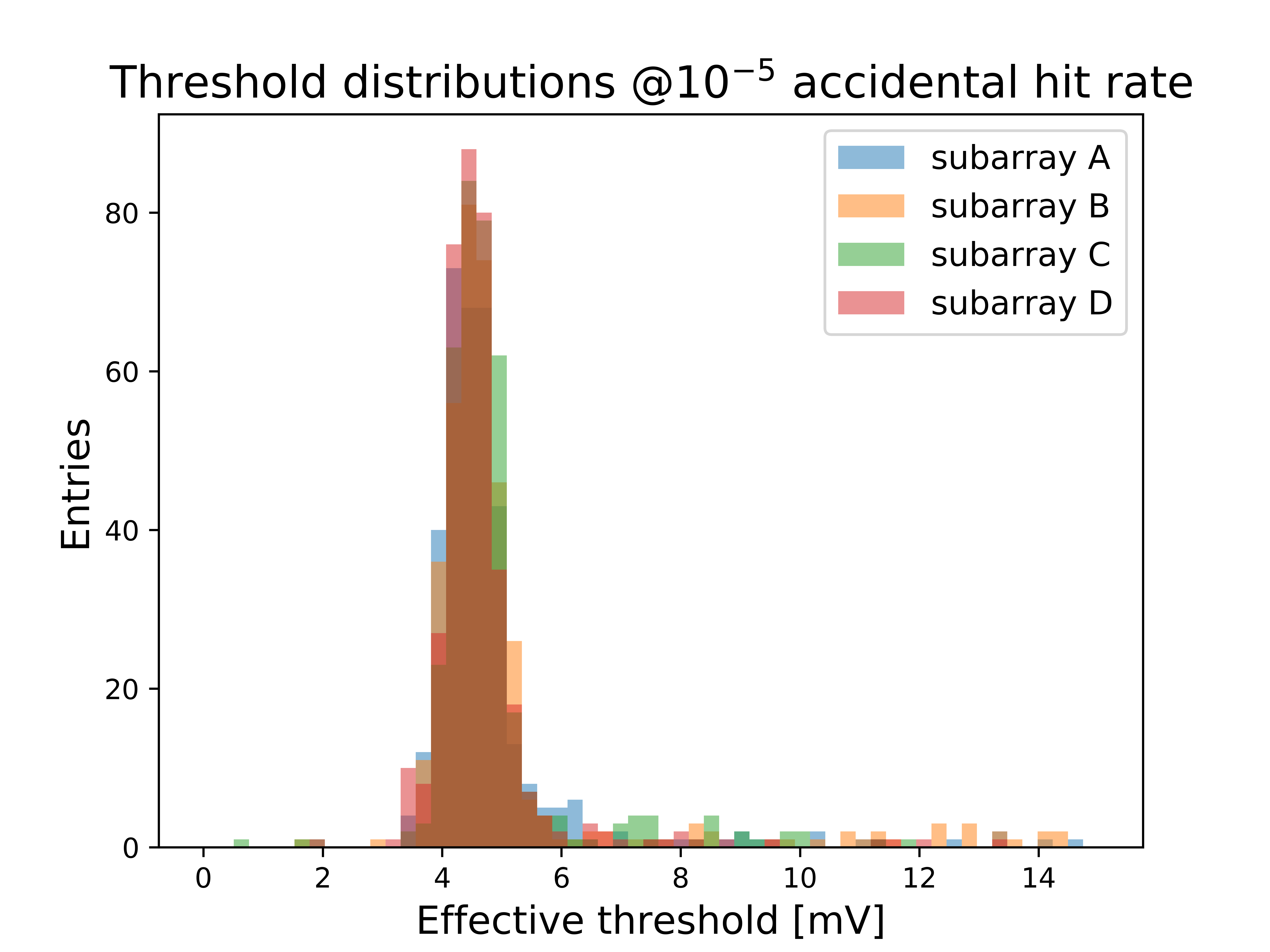}
\caption{\label{fig:noise_thresholds}Sensor functional parameters as measured before the installation of the 2016 PXL detector. Left: sensor sub-array total noise distributions. Right: effective sub-array discriminator threshold distributions corresponding to a 10$^{-5}$ accidental hit rate equal.}
\end{figure}

The test results for all the ladders installed on the detector operated in the 2016 Run are summarized in Figure \ref{fig:noise_thresholds}. The left panel shows the distributions of the total noise as measured on the four sub-arrays forming the 400 detector sensors. The overall noise is distributed around 1.15~mV with a $\sigma$~=~0.17~mV. The Fixed Pattern Noise component contributes for around 1/10 of the total noise. The distributions of the effective sub-array discriminator thresholds used to achieve a 10$^{-5}$ accidental hit rate are shown in the right panel. Around the 90\% of the sub-array thresholds are distributed around 4.6~mV with a $\sigma$~=~0.5~mV.

\begin{figure}[h!]
\centering 
\includegraphics[width=.65\textwidth]{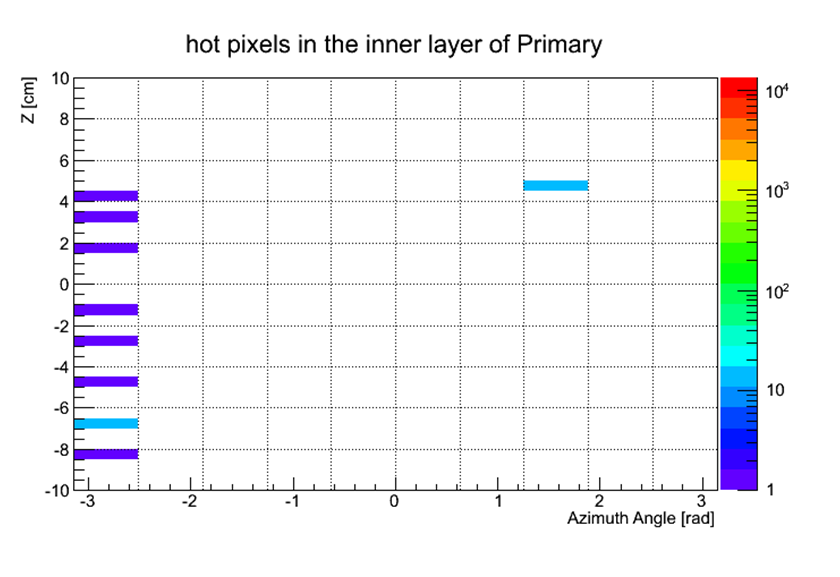}
\caption{\label{fig:innerhot}Inner layer $z$-$\phi$ distribution of the number unresponsive pixels per sub-array on the 2016 Run PXL detector, as measured before the installation.}
\end{figure}

\begin{figure}[h!]
\centering 
\includegraphics[width=.95\textwidth]{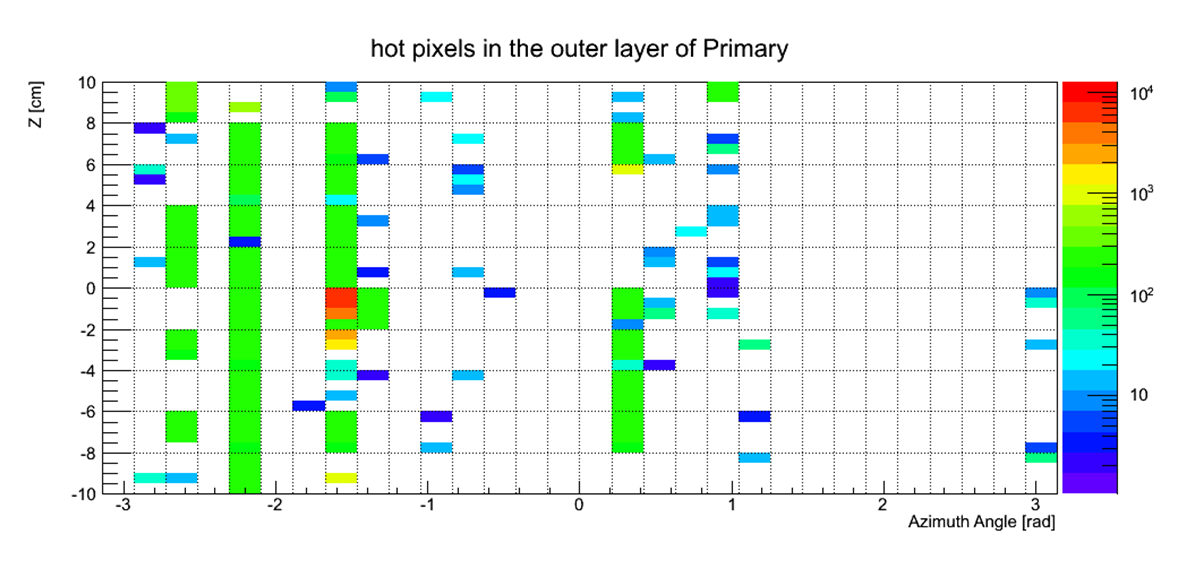}
\caption{\label{fig:outerhot}Outer layer $z$-$\phi$ distribution of the number unresponsive pixels per sub-array on the 2016 Run PXL detector, as measured before the installation. Only \emph{hot} pixels (always above threshold) are considered.}
\end{figure}

The distribution of malfunctioning pixels per sub-array as measured before the installation of the 2016 Run detector is shown in Figures \ref{fig:innerhot} and \ref{fig:outerhot}, projected on the detector $z$-$\phi$ plane. Only \emph{hot pixels} (always above threshold) are visible in the histograms. A few sensors present a non-negligible number of unresponsive pixels, due to the fact that slightly lower quality ladders were used for the last refurbishment of the detector.

\subsection{Metrology survey} \label{sec:metrology}
\begin{figure}[htbp]
\centering 
\includegraphics[width=.38\textwidth, trim=1cm 0 1cm 0]{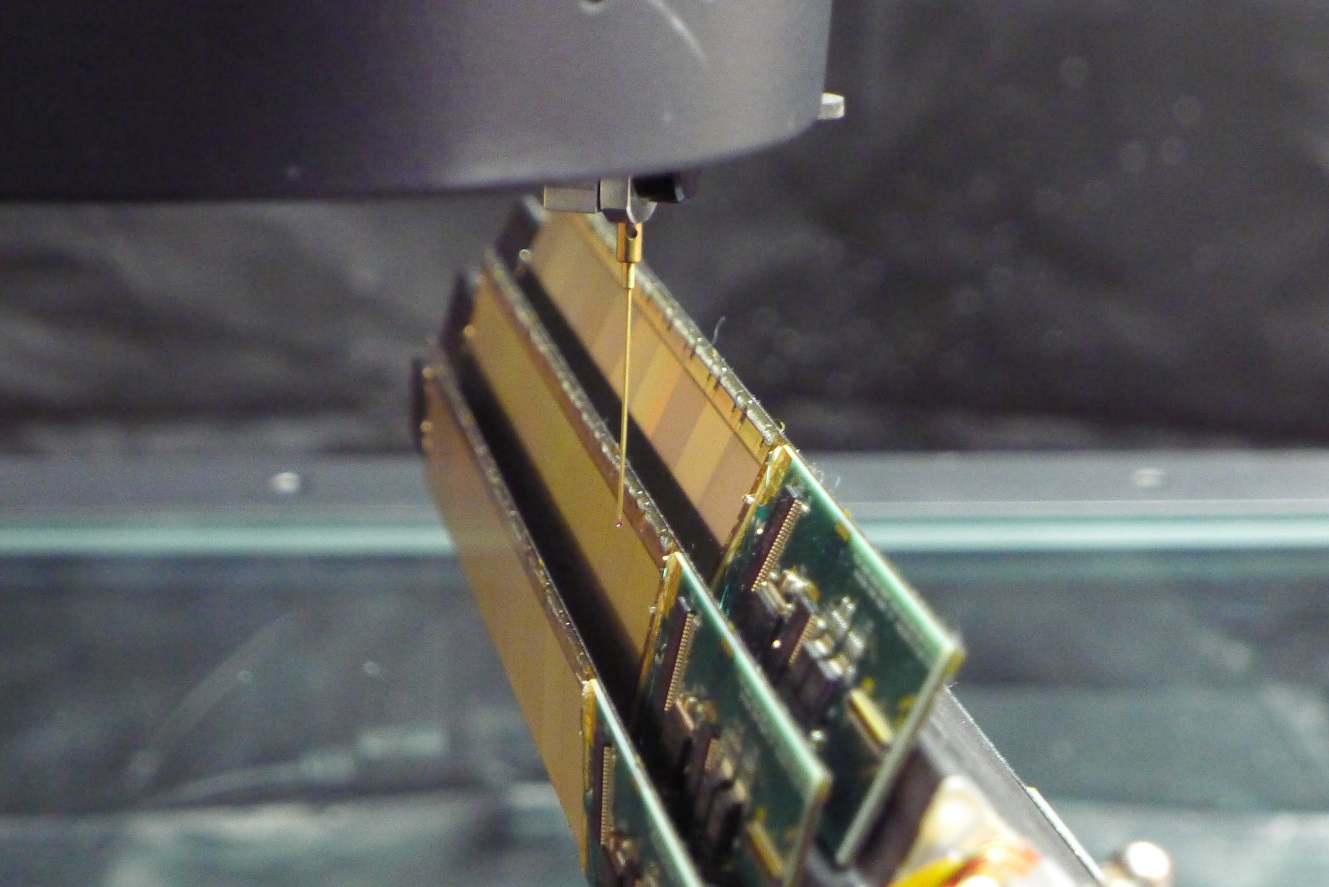}
\qquad
\includegraphics[width=.48\textwidth]{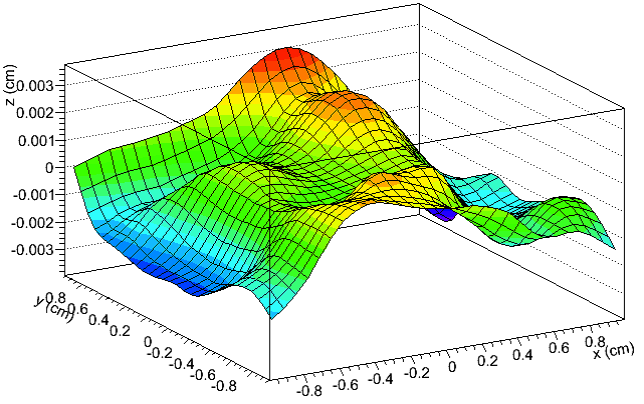}
\caption{\label{fig:metrology}Left: 3D-measurement of the locations of all pixels on a sector. Right: example of sensor surface profile.}
\end{figure}

An accurate control of the sensitive element position is crucial to achieve the required position resolution. Fully assembled sectors are surveyed in a CMM with a programmed automatic procedure: the position of two fiducial lithography markers in each chip is measured with optical head, with a resolution of 3~$\mu$m in xy and 50~$\mu$m in z. The sensor surface profile is then measured with an 11-by-11 point pattern using a \emph{Feather Probe}\footnote{\href{https://www.ogpnet.com/north-america/accessories/micro-probe-options/feather-probe/index?page=41}{https://www.ogpnet.com/north-america/accessories/micro-probe-options/feather-probe/index?page=41}} with 3~$\mu$m resolution in the three coordinates. The sector survey setup is shown in Figure \ref{fig:metrology}, left picture. This touch-probe permits picking up over hung surfaces to map the 3-dimensional locations of all pixels on a sector with respect to three sector tooling balls (see example in Figure \ref{fig:metrology}, right panel) mounted on the sector tube.

The PXL sensor surface profile measured through the survey shows deviations of $\pm$30~$\mu$m from the nominal ladder location. The chip-to-chip surface position deviations are even larger along the ladder surface. These deviations are larger than the measured hit position resolution. In order to take into account the deviations, the mapped detector geometry is included in the reconstruction software.
Five sectors are mounted in a half shell to form a detector-half. The detector-half is installed on a truncated PXL support tube with kinematic mounts, reproducing the final PXL location in the STAR detector, and surveyed in a second precision CMM. The position of the sector tooling balls is measured with touch probe relative to the kinematic mount coordinates to form completely mapped stable units, with the pixel positions known to an accuracy of approximately 10~$\mu$m. Since the shells are supported the same way in the CMM and in the STAR installation, the relative pixel position mapping remains undisturbed. 

\section{Detector operations}\label{sec:operations}

\subsection{Integration with STAR}
\subsubsection{Slow control}\label{sec:slowcontrol}
To control and monitor the RDO boards we use the FTDI USB interface chip FT2232H as described above. The chip is programmed to provide two sets of asynchronous FIFO interfaces to the FPGA. Each FIFO interface consists of 8 bits of data and various FIFO control lines. The FPGA can read and write to the data lines by setting the appropriate FIFO control lines. On the firmware side, we use one of the FIFO interfaces for communication over USB, while the second FIFO interface is used to reset various internal FPGA state machines associated with the USB communications. We developed a custom communications protocol for this project, which presents 2 firmware FIFOs to the USB interface chip, one ``command FIFO'' that is used for communication from the PC (via USB) to the FPGA, and one ``response FIFO'' for data from the FPGA to the PC. The USB interface protocol from the PC to the FPGA consists of 2 kinds of packets: a ``write'' packet and a ``read'' packet. Each packet starts with a 32 bit ``header'' word that identifies the packet as either ``read'' or ``write'', an ``address'' for this transaction, and defines the number of words following the header which contain the actual ``payload'' of the packet. A ``write'' packet will result in some parameter in the firmware being changed (through a set of memory mapped firmware registers), while a ``read'' packet will result in data being written to the response FIFO, which can then subsequently be read by the USB interface. To write to FPGA memory, we defined a set of ``indirect'' registers that need to be written to in sequence: an ``address'' register to define the memory address to write to or read from, and a ``data'' register to write to or read from this address. The address register auto-increments, so that multiple writes or reads to this memory can be performed without having to set the address of each transaction. For ``memory read'' transactions an additional ``count'' register allows multiple reads from the FPGA memory in response to one USB ``read'' transaction.

On the PC side we chose the open-source library ``libftdi'' which provides a C/C++ interface to the FT2232H chip based on ``libusb'' for both the Windows and Linux operating systems. In order to map the functions provided by libftdi to our custom USB protocol, we wrote a C++ library containing functions to read and write registers, as well as write to and read from memory. To facilitate easy prototyping and scripting via the programming and scripting language ``Python'', we mapped the functions of this library into equivalent Python commands via the Python extension API. An added benefit of this port is the possibility of writing simple graphical user interfaces (GUIs) to the scripts for operator control and monitoring of the PXL electronics using various GUI APIs provided within the Python environment.

The STAR experiment standardized the detector control and monitoring (``Slow Controls'') to the software environment ``EPICS'' (Experimental Physics and Industrial Control System) \cite{epics}. EPICS is a set of Open Source software tools, libraries, and applications developed to create a distributed real-time control system for particle accelerators, telescopes and large scientific experiments, typically consisting of tens to hundreds of computers, networked together to allow communication, control, and monitoring of the various instruments in such a system from a central control room, or remotely over the internet. The basics architecture of EPICS is a client/server model that uses ``publish/subscribe'' techniques to communicate between the various instruments and the computers; the servers in this model are called ``Input/Output Controllers'' (IOCs) and attached to instruments performing real-world I/O and local control tasks, and publish this information (so called ``Process Variables'', or PV's) to clients using the ``Channel Access'' network protocol. The EPICS system provides drivers and software for these IOCs, as well as various libraries for building control and monitoring software interfaces. Originally, all the IOCs of the EPICS system were single board computers running the vxWorks real-time operating system and were installed in a VME chassis. In modern EPICS systems, however, an IOC can also be an embedded microcontroller, a rack-mount server, or even a desktop PC, running other operating systems, but always attached to hardware devices performing input/output operations. For devices that do not have drivers and libraries in the EPICS environment, a host-based IOC called ``soft-IOC'' can be developed, which translates the EPICS environment into the protocol the associated hardware supports. Most operations in EPICS are driven from Graphical User Interfaces (GUIs) created with EPICS supplied standalone GUI packages such as MEDM (Motif Extensible Display Manager), which is the GUI package of choice in STAR.

In PXL, we use EPICS to interact with various components of the detector: monitoring and control of the  power supplies for the PXL sensors and the power supply in the electronics crate which powers the RDOs;  monitoring of various electronics parameters read from the RDO boards (via USB), and the monitoring and control of the cooling system. All of these systems are controlled and/or monitored via soft-IOCs in the STAR EPICS system.

Since the USB interface to the RDO boards is intimately related to the configuration of the PXL electronics and needed various specialized interfaces, we decided not to use EPICS for the control of the electronics, but only for monitoring, alarming and archiving of the various parameters read over the USB. The Python interface scripts described above used an open-core Python interface library to EPICS to update the PV's in a soft-IOC in EPICS, which in turn was used to create alarms and to provide archiving for the monitored parameters.

To power the detector, three TDC-Lambda ``GEN8-180-LANGenesys'' DC power supplies are used. Each power supply is capable of delivering 180 A at up to 8 V with a 265 VAC input. Two of these supplies are used to deliver the 4 V power required by the ladders, one supply each for 5 sectors; the third supply delivers the 6 V power for the MTBs on all sectors. These power supplies provide manual control and monitoring from their front panel, as well as remote control and monitoring over Ethernet via an LXI-compliant protocol. LXI (``LAN Extension for Instrumentation'') is a standard adopted by many test and measurement companies to provide network control and monitoring to their instruments. The ``SCPI'' (Standard Commands for Programmable Instrument'') command set was used to create a soft-IOC for EPICS, based on development work done within the EPICS community for similar supplies. A set of MEDM based GUIs allows the control and monitoring of these supplies. A database interface to EPICS is used to archive voltage and current measurements from the supplies, while EPICS itself provides means to set software alarms for current or voltage values that fall outside an expected range.

The power supply in the crate that houses the RDO boards is a standard VME crate power supply from ``Wiener, Plein \& Baus, Corp'' and provides monitoring and control to the network via the ``Simple Network Management Protocol'' (SNMP). A soft-IOC developed for a similar supply in another detector of STAR was used to interface this supply to the EPICS system. Just as for the Lambda supplies, an MEDM GUI was developed for monitoring and control in the STAR control room, while the voltage and current values are archived in the EPICS database.

\subsubsection{Trigger system}
Trigger information is delivered to the PXL electronics boards from the the STAR Trigger subsystem via the ``Trigger and Clock Distribution'' (TCD) board. With each trigger decision a 4-bit trigger command, 4-bit DAQ command, and a unique 12-bit token is distributed to all sub-detectors that are involved in a specific trigger. The trigger command determines the kind of action the sub-detectors are supposed to take (allowing for up to 15 different actions, since the value ``0'' is reserved to indicate an ``idle'' condition). The DAQ command is forwarded un-altered from the detectors to DAQ for interpretation. The token allows for several triggers to be handled simultaneously in STAR, but each token can only be used once until the associated event has been built and forwarded for storage in the RHIC mass-storage system. 

The physical connection between the TCD and the PXL electronics is accomplished over a 30~m 20-conductor shielded twisted-pair cable between the TCD crate on the electronics platform and the crate that contains the PXL electronics. Eight pairs of this cable distribute ``Pseudo-emitter coupled logic'' (PECL) signals from the TCD to the electronics. One pair returns a busy signal to the TCD, while another pair is reserved for return of a status, but not used by the PXL system. The signals from the TCD to the PXL electronics are:
\begin{itemize}
\item A clock synchronous with the RHIC accelerator bunch crossing clock (typically around 9.38~MHz for 200~GeV Au-Au)
\item A data clock for distributing the trigger and DAQ commands as well as the token.
\item 4 data bits
\item 2 additional detector specific clocks (not used by the PXL electronics)
\end{itemize}
As mentioned above, the trigger data to be distributed consists of a total of 20 bits. In order to distribute these 20 bits within one RHIC clock period over the 4 data bit lines, the data clock is running at 5 times the frequency of the RHIC clock.

On the PXL electronics side, the TCD cable terminates in a separate board in the PXL electronics crate that distributes all of the downstream signals to the 10 RDO boards, receives and combines with an OR operator the busy signals from each RDO board, and transmits a PXL global busy back the to the TCD. In order to protect the time from the trigger decision until the trigger busy is returned to the TCD, the TCD logic generates an internal busy which is returned to the trigger system.

The hardware based, so-called ``Level-0'' (L0) trigger is created and distributed by the trigger system within about 1.5~$\mu$s; an additional 250~ns is due to cable lengths and signal distribution, so L0 triggers arrive at the PXL RDO boards around 1.75~$\mu$s after the collision that caused the L0 trigger decision. One of the trigger commands is used to signal this trigger decision to the electronics. Online processing of the events results in two additional levels of trigger, that allow STAR to accept or abort triggered events later in the readout chain. Two trigger commands are used for these higher level triggers and indicate an ``accept'' or ``abort'' to the detectors. In the PXL RDO these trigger commands are received, but result in no specific action in the RDO. These received triggers and their associated DAQ-command and token are passed on to the DAQ receivers for further processing in the DAQ system. 

In the RDO firmware the number of RHIC periods received from the TCD is counted in a 32-bit register and passed along with the event data for L0 triggers. One of the trigger commands is used to reset this counter. Another trigger command is used for a general reset of all RDO logic associated with the TCD interface. All other trigger commands are received and passed along with the event data, but ignored by the internal RDO logic.

\subsubsection{Data Acquisition System (DAQ)}
The interface to the DAQ system is provided by the ALICE DDL as described above, consisting of a plugin board for the RDO board called ``Source Interface Unit'' (SIU), a bi-directional optical fiber, and a ``Readout Receiver Board'' (RORC) that plugs into a PCI-X port on a DAQ PC. The SIU interface to the FPGA firmware presents a 32bit bi-directional data port and various control signals that determine the kind of data being transferred as well as the direction. A set of feedback signals allows throttling of the data flow in either direction, if resource starvation demands. Based on the 32bit data interface, the event format from the PXL RDOs to the DAQ computers consists of a sequence of 32bit words from each of the 10 RDO boards subdivided into various sections: a 16 word header (containing various housekeeping data), a word defining the number of words in the subsequent ``Hit Block'', a variable-length block of ``Hit Addresses'', a separator word to indicate the end of the hit block as well as the beginning of the subsequent block, a variable-length block of Trigger (TCD) data, a word containing the ``Cyclic Redundancy Code'' (CRC) of the preceding words, and finally an ``Ender'' word terminating the data sequence. As described above in the trigger interface section, the trigger data contains all of the trigger data received from the TCD during time it takes to read out an event, including the trigger data that initiated the readout. The DAQ system interprets the trigger commands contained in these data to decide whether to abort or accept this and previous events (since a higher level trigger can take some time to arrive at the RDO boards, at which time the affected event might already have been sent to the DAQ system). To allow the DAQ system to finish events in a timely manner, the RDO boards are sending ``trigger only'' events to the DAQ system, if trigger data has been accumulating in the RDO boards, when no new event readouts are triggered within a pre-defined amount of time (typically one millisecond). 

\subsection{Scripted operations}
The PXL detector is controlled and monitored through a multi-threading Python-scripted software integrated with the general STAR EPICS framework and with the PXL specific read-out and power systems, as described in Section \ref{sec:slowcontrol}. The operations are controlled through a ``\emph{TKinter}'' module-based python GUI. 

\begin{figure}[h!]
\centering 
\includegraphics[width=.95\textwidth]{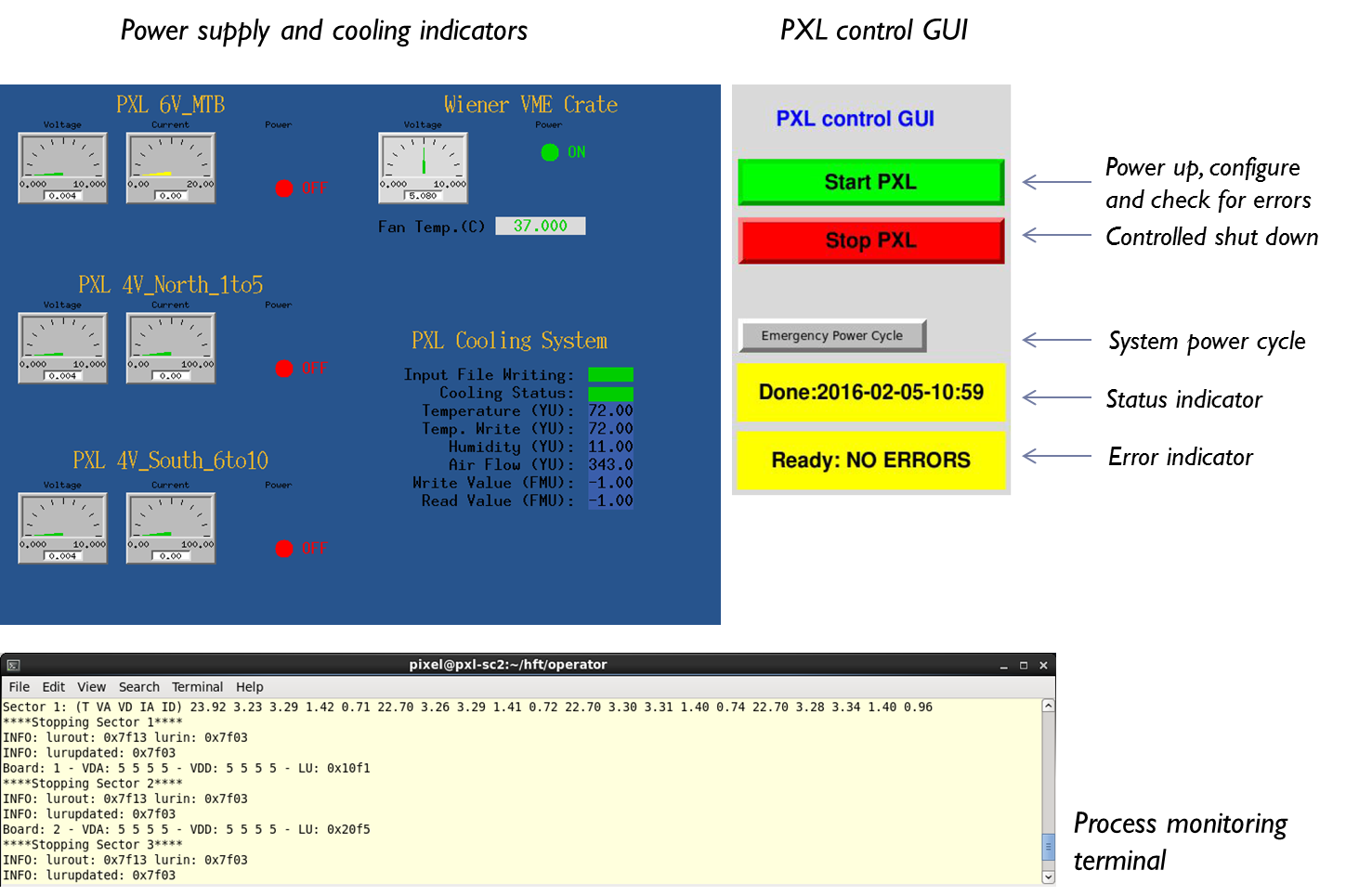}
\caption{\label{fig:GUI}PXL detector operations control and monitoring GUI.}
\end{figure}

The simple GUI layout shown in Figure \ref{fig:GUI} simplifies the detector operator actions to a small set of safe operations:
\begin{itemize}
\item \emph{Start PXL} - In order to operate the detector in the common STAR data taking, the ``\emph{Start PXL}'' command enables the power supplies and executes serially the following operations on the 10 sectors:
\begin{enumerate}
\item set the proper I/O delay to synchronize the sensor input signals
\item enable the power circuits on the MTBs
\item set the proper output voltages and overcurrent protection thresholds 
\item send the proper JTAG configuration and activate the sensors
\item enable the latch-up protection circuit
\end{enumerate} 
After completing the described procedure, the detector is ready to be included in the STAR detector cluster and to take data.

\item \emph{Stop PXL} - In order to stop the detector operations, the ``\emph{Stop PXL}'' command removes the power in a controlled fashion and disables the power sources.

\item \emph{Emergency Power Cycle} - In case of Single Event Upset (SEU) experienced by the RDO board FPGAs resulting in a sudden readout firmware misbehavior, the ``\emph{Emergency Power Cycle}'' command operates a complete power cycle and reconfiguration of detector and RDO electronics. The procedure triggers a fast FPGA configuration from the on-board parallel flash memory and clears the SEU-induced error.
\end{itemize}
A series of periodic asynchronous tasks is also automatically executed in the background by the multi-threaded software: 
\begin{itemize}
\item 15 minute cycle reconfiguration of all the sensors to clear possible errors generated by SEU events
\item 1 minute cycle monitoring of the following detector functional parameters:  analog and digital currents and voltages, detected latch-up events, sensor temperature, sensor frame synchronization errors. The monitored parameters are archived into the STAR EPICS database and made accessible via web interface.
\end{itemize}
The detector control software also displays the executed commands and the system response in real-time, and logs them into a text file available for later consultation.  

\subsection{Detector calibration}

\begin{figure}[h!]
\centering 
\includegraphics[width=.95\textwidth]{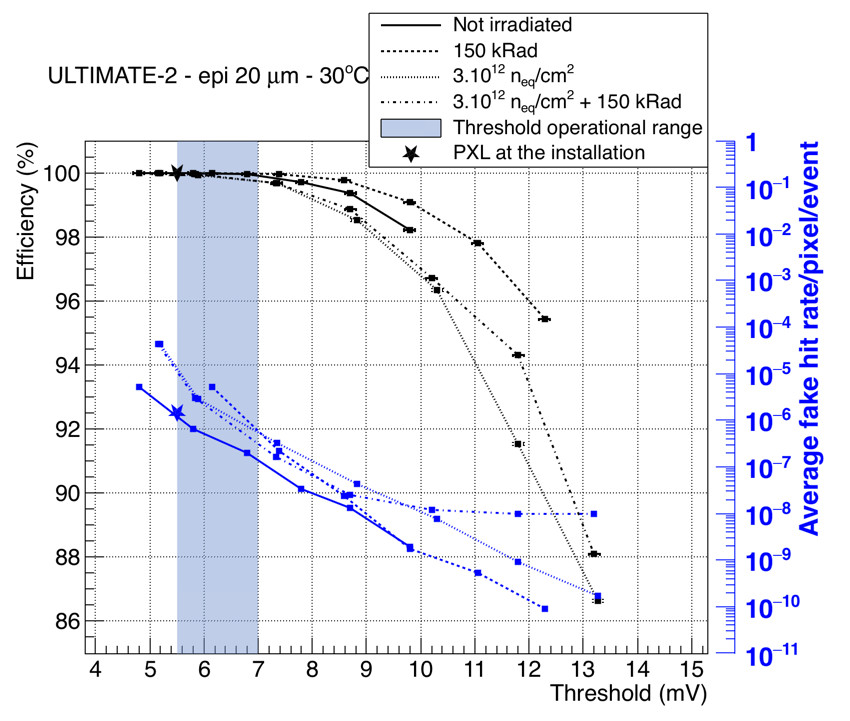}
\caption{\label{fig:PXL_thresholds}PXL detector operating thresholds: the chosen thresholds guarantee an accidental hit rate of 1.5$\cdot$10$^{-6}$ and a detection efficiency larger than 99.5\%. On average, the applied threshold at the installation is equal to 5.48~mV, corresponding to about 4.12~$\sigma_{noise}$. Irradiation tests demonstrated that the threshold operational margin (blue area) allows for a similar efficiency throughout the detector lifetime}
\end{figure}

The calibration of the PXL detector is designed to maximize the detection efficiency while minimizing the accidental hit rate. The discriminator thresholds of each sensor are adjusted in order to achieve an accidental hit rate of 1.5$\cdot$10$^{-6}$, which corresponds to about one pixel above threshold per sensor per event. 
Once installed on a ladder, each sensor is individually powered and fully characterized in controlled temperature and light conditions. The discriminator transfer function characteristics are initially measured along with noise in test mode, i.e. through a full matrix non-zero suppressed readout. The sensor is then operated in fast-readout mode (with zero-suppression), and the discriminator thresholds are adjusted to provide the desired accidental hit rate. This preliminary calibration settings are stored in configuration files for later loading via JTAG protocol.
After the installation of the PXL detector in STAR, the discriminator thresholds of each sensor are optimized by scanning a narrow range around the preliminary calibration settings while operating the rest of the detector in running condition, in absence of beam in the machine. On average, the applied threshold is equal to 5.48~mV, corresponding to about 4.12~$\sigma_{noise}$, where $\sigma_{noise}$ is the average total noise (see Figure \ref{fig:PXL_thresholds}). The irradiation tests on Ultimate sensors demonstrated that the threshold operational margin allows for a detection efficiency larger than 99.5\%, even after the exposure to an ionizing radiation dose of 150~kRad and non-ionizing radiation dose of 2$\cdot$10$^{12}$~1MeV~n$_{eq}$/cm$^2$. 

In the case of any permanent change in the current drawn by a ladder, which may result from a latch-up induced damage or from disabling an unresponsive sensor, the over-current protection thresholds are re-calibrated and set to 80~mA above the operating current.

During the scheduled biweekly $\sim$8 hour stop of the RHIC operations, or whenever a sufficient time period was made available for dedicated detector operations in absence of beam, a series of extensive calibrations was repeated to optimize the detector configuration and the power system settings:
\begin{itemize}
\item Repeat the sensor discriminator threshold calibration as described above
\item Mask off the pixels/columns/sub-arrays/sensors unresponsive to threshold variations 
\item Adjust the digital and analog voltage output for each ladder in order to compensate for the voltage drop, as calculated on the basis of the corresponding operating current
\end{itemize}

The sensor noise was not directly evaluated after the detector installation due to technical limitations. The simplified set of measurements performed to first calibrate the detector confirmed the performance measured before the installation.  

\subsection{Detector safety}
The PXL detector safety is ensured by a series of 4 interlock signals connected in series. The input signals are generated by: 
\begin{itemize}
\item the global STAR interlock: triggered by any unsafe STAR equipment condition
\item a temperature controller monitoring the air temperature at the chiller system outlet: triggered when the supplied air temperature rises above $\sim$26$^{\circ}$C
\item two airflow probes, placed in the two detector-half air duct inlets: triggered when the supplied airflow falls below half of the nominal value,  corresponding to about 5~m/s air velocity
\end{itemize}
Any of these interlock conditions triggers an immediate shut-down of the PXL power supplies.

\section{Detector performance}\label{sec:performance}
The PXL detector was installed and operated for the first time in STAR in January 2014.
After being successfully commissioned, it collected $\sim$1.2 Billion minimum-bias Au+Au events at $\sqrt{s_{NN}}$~=~200~GeV during the 2014 Run. In the 2015 Run it collected $\sim$1 Billion p+p and $\sim$0.6 Billion p+Au events. The data sample was completed in the 2016 Run, with $\sim$2 Billion Au+Au and $\sim$0.3 Billion d+Au events.

In the final year of operations, after constant improvements of the RHIC and STAR performance, the PXL detector took data at a typical trigger rate of 0.8-1~kHz with a dead time of $\sim$6\%, handling a maximum number of pixels above threshold per sensor per event equal to 1000 (100) on the inner (outer) layer, corresponding to a maximum occupancy of $\sim$0.1\% ($\sim$0.01\%). During data taking operations, each sector was reconfigured after a latch-up event at an average rate of 12/hr. The full detector recovery time from detection of the latch-up event to the reconfigured sector ready for data taking was $<$~1~s.
At the beginning of each Run, a PXL detector copy featuring an active channel fraction larger than 99\% was installed in STAR. The noise measured on the installed detector in absence of beam activity appears to be roughly 20\% lower than the noise measured during sensor probe testing. We attribute the difference to the lower quality connections to power and ground provided by the probing pins as well as the lack of properly placed capacitor bypassing of the sensors.

The SSD and IST subsystems have been operating in stable conditions for most of the first two years of operations, accounting for 80\% and 95\% active channels respectively.

The PXL performance as measured from the data collected in the 2014 Run are reported in this section. 

\subsection{Efficiency}
\begin{figure}[h!]
\centering 
\includegraphics[width=.5\textwidth]{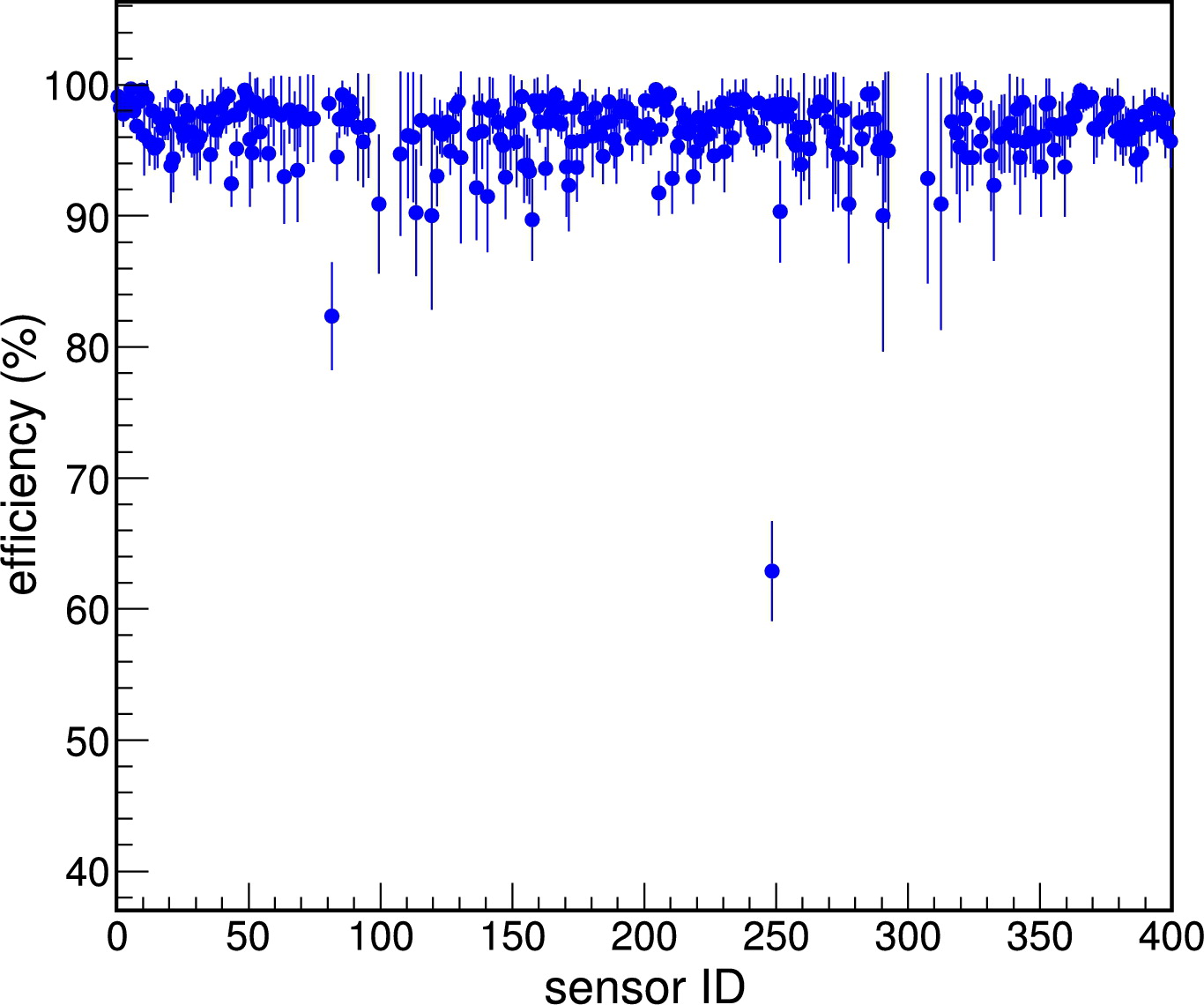}
\caption{\label{fig:efficiency}PXL sensor efficiency as a function of the sensor ID, measured in cosmic ray data with an unrefined configuration of the sensor discriminator thresholds and a preliminary tracking software version. The average efficiency over all sensors is 97.2\%. The gaps around sensor IDs 80 and 300 correspond to the horizontal direction.}
\end{figure}

The efficiency of the PXL detector was measured with cosmic ray data taken with zero magnetic field by comparing the position of a measured hit on a sensor with a straight line fit from three other hits \cite{efficiency}. The results are shown in Figure \ref{fig:efficiency}. The data were taken before the 2014 Run beam operations with an unrefined detector configuration and reconstructed with a preliminary tracking software version. The average efficiency, over all sensors, was measured to be 97.2\%. The gaps around sensor IDs 80 and 300 due to the dearth of cosmic rays in the horizontal direction.

\subsection{Alignment}
\begin{figure}[h!]
\centering 
\includegraphics[width=.98\textwidth]{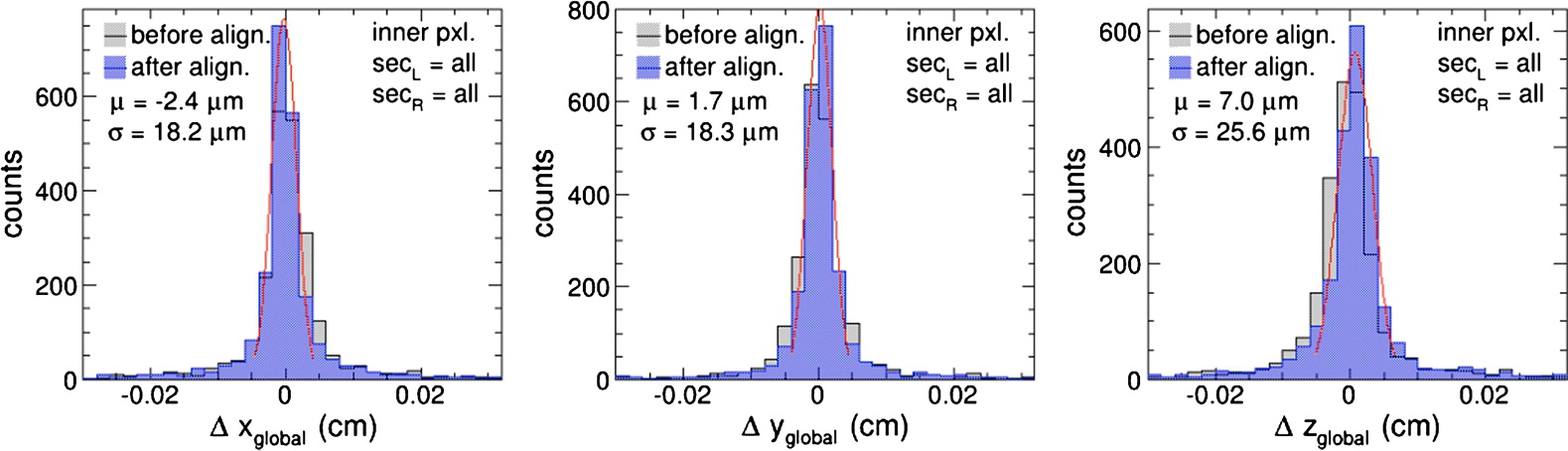}
\caption{\label{fig:alignment}PXL hit residual to cosmic track before and after PXL sector alignment. The Gaussian fit to the residuals after alignment shows a $\sigma\leq$~25~$\mu$m.}
\end{figure}

In order to fully exploit the potential of the HFT, the detector geometry was surveyed and measurements were made to determine any deviation of the detector elements from their designed positions, as described in Section \ref{sec:metrology}. After the installation, the different parts of the HFT were aligned using cosmic ray data. Figure \ref{fig:alignment} shows PXL hit residuals compared to cosmic track projections before and after PXL sector alignment. The Gaussian fit to the residuals after alignment shows a $\sigma\leq$~25~$\mu$m, which meets the design goals considering both hit errors and multiple Coulomb scattering \cite{bib_survey}. 
The detector stability required during the initial alignment period is the same as the stability required during data taking. The pixels should not deviate from their equilibrium positions by more than 6~$\mu$m FWHM.  
Primary track alignment is done during the periods either before the run, after the run when straight (magnet off) cosmic ray muon tracks that traverse the full detector (4 hits) can be used. During the run periods when RHIC is undergoing injection or there are beam off periods, additional cosmic ray data is taken s a check on the alignment throughout the run. With one minor exception, the locations of the detector ladders was unchanged over the duration of the yearly run.
As has been explained throughout the detector mechanics description, great efforts have been taken to maintain detector positional stability throughout the run and this has been quite successful. We did not experience significant excursions from nominal airflow speed or temperature and new alignments were not required due to environmental factors.

\subsection{DCA pointing resolution}
\begin{figure}[h!]
\centering
\includegraphics[width=.90\textwidth]{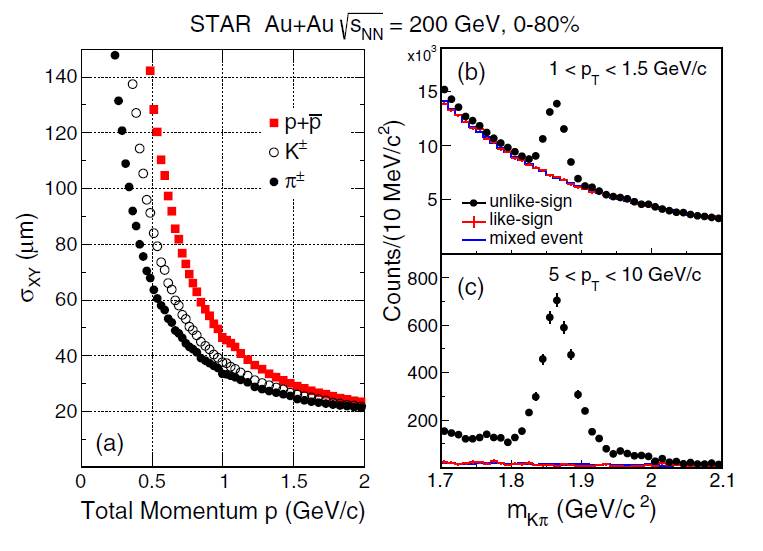}
\caption{\label{fig:performance}Left: Track pointing resolution in the azimuthal direction as a function of the particle momentum (a), measured for the overall detector. Similar resolution has been measured for the track pointing resolution along the beam direction. Right: $D^0 \rightarrow K \pi$ production in 0\%-80\% centrality $\sqrt{s_{NN}}$~=~200~GeV~Au+Au collisions, for 1~$<$~p$_T$~$<$~1.5~GeV/c (b) and 5~$<$~p$_T$~$<$~10~GeV/c (c). The reconstructed $D^0$ signal (solid data points) are compared with the combinatorial
background distributions (red crosses and blue lines), estimated with the like-sign pairs and the mixed event unlike-sign techniques, respectively) \cite{D0}.}
\end{figure}

The HFT internal alignment was conducted with a combination of high precision survey measurements for sector internal positions and cosmic ray tracks in zero magnetic field for inter-sector alignment. The analysis of Au+Au data collected in 2014 Run demonstrated that the track pointing resolution of the HFT system meets the design requirements, achieving $\sim$46~$\mu$m for 750~MeV/c kaons for the 2 sectors equipped with aluminum cables on the inner layer, and better than 30~$\mu$m for particle momenta $\geq$~1~GeV/c. The track pointing resolution in the azimuthal direction as a function of the particle momentum, as measured for the overall detector, is shown in Figure~\ref{fig:performance}, left panel. This performance enabled the study of \emph{D}-meson production with a high significance signal. The measurement of the $D^0 \rightarrow K \pi$ production in $\sqrt{s_{NN}}$~=~200~GeV~Au+Au collisions for 1~$<$~p$_T$~$<$~1.5~GeV/c (b) and 5~$<$~p$_T$~$<$~10~GeV/c (c) is shown in Figure~\ref{fig:performance}, right panel. The projected significance for the entire 2014 Run data sample reaches 220. The first measurement of the elliptic anisotropy (\emph{v2}) of the charm meson D$^0$ at midrapidity ($|\eta|\leq$~1) in Au+Au collisions at $\sqrt{s_{NN}}$~=~200~GeV has been published \cite{D0}.
%
%

\section{Lessons learned}\label{sec:lessons}
PXL detector design choices and construction procedures provide a wealth of information that can be useful for future vertex detectors. A brief description of selected lessons learned is presented below.
\paragraph{Detector construction} Efficient assembly of PXL ladders required selecting only high quality sensors that had been diced and thinned to 50~$\mu$m. Development of a probe test setup required the design of a dedicated vacuum chuck that would allow for individual loading of up to 20 thin sensors for automated testing. Testing of thin chips, which typically curve at this reduced thickness, proved to be challenging and the default probe pin design delivered by the probe card vendor had to be optimized to allow for additional overdrive distance to improve contact reliability with a non-flat surface. The changes were the increase in beam length of the probes along with the probe taper which had the functional effect of reducing the spring constant of the probe and giving longer travel for equivalent probe force. The probe card was equipped with readout electronics necessary to perform full sensor characterization, including noise and operating threshold measurements, as well as the full speed readout at 160~MHz. The testing functionality built into the sensors was crucial at this stage. The overall yield of the detector grade sensors during the PXL detector production varied between 46\% and 60\%.
 
The construction of PXL ladders relied on custom made vacuum fixtures that provided high precision alignment of individual ladder components. The sensors were manually placed, butted together, and glued to the readout flex cable using a low elastic modulus acrylic adhesive, preventing excessive position warping and detector damage from different thermal expansion characteristics. Sensor alignment on the ladder was significantly enhanced by the use of Deep Reactive Ion Etching (DRIE) during wafer production to create dicing trenches. Sensors were connected to the flex readout cable via standard wire bonding and the wires were encapsulated for protection. 

The production of the double-sided aluminum flex cables encountered several technical difficulties and was delayed. For this reason, the first PXL detector constructed featured only two inner ladders equipped with aluminum-conductor flex cables. The remaining 38 ladders were assembled using a copper-conductor backup alternative. The use of copper instead of aluminum effectively doubles the material budget of the PXL flex cable and translates to an increase in the ladder radiation length from approximately 0.4\% to 0.5\%. 

\paragraph{Engineering Run}
A crucial milestone in the PXL detector development was the engineering run performed with a prototype detector inserted into STAR for several weeks of the 2013 RHIC run \cite{Michal_EngRun}. The prototype detector featured full mechanical support structure and was equipped with three sectors and the associated readout electronics. The initial production of ladders for the engineering run detector revealed several issues that lead to short circuits in the assembled ladders. The issues were resolved by extending the adhesive tape layer beyond the sensor footprint and by adding a solder mask layer to the flex cable corresponding to 0.0075\% radiation length (this layer was initially removed to minimize the material budget of the assembly). The assembly of sectors for the engineering run revealed a mechanical interference between driver boards located at the end of inner ladders in neighboring sectors. As a consequence, the engineering run detector featured full sectors separated by at least one empty sector (a sector tube without any ladders attached). Ultimately, the geometry of the sector tube required modifications and the radius of the inner layer increased from the original 2.5 cm to 2.8 cm. The engineering run experience led also to optimization of the MTB. The new board features remote monitoring of the individual ladder current consumption (analog and digital) and remote control of the over-current protection threshold and the ladder power supply voltage.

\paragraph{Latch-up induced sensor damage}
During the commissioning phase of the PXL detector with 14.5~GeV~Au+Au collisions, the detector started to exhibit performance changes that indicated radiation-related sensor damage. The damage continued to accumulate into the first two weeks of the 200~GeV~Au+Au run until a set of operational methods was applied, which effectively limited further damage to the sensors. The observed damage appeared to be latch-up related and took on many different forms: increased digital current consumption, damage to pixel columns, loss of full or partial pixel sub-arrays, etc.
A total of 16 of the 400 sensors in the PXL detector were damaged, which corresponds to a loss of 14\% of the active surface on the inner layer and 1\% on the outer layer.  
In the set of operational methods applied to limit the observed sensor damage, the most important modification was the reduction of the over-current protection threshold. The initial threshold of 400~mA above the measured operating digital current on each ladder was decreased to 120~mA. In addition, the PXL detector was only turned on for data taking when the collision rate was below a certain value and all the sensors were power-cycled and reconfigured every 15 minutes. With all these procedures in place, only two more sensors exhibited additional damage in the remaining three months of the run. 


The failure mechanism has been extensively studied through a test campaign carried out in Fall 2014 at the 88” Cyclotron BASE Facility at the Lawrence Berkeley National Laboratory (LBNL) where existing PXL ladders and sensors have been exposed to heavy ion and proton beam irradiation. This test allowed the characterization of the damage mechanism and for the definition of a safe operation procedure to set up the power supply current threshold settings for the subsequent data taking periods. The sensors tested included 50~$\mu$m and 700~$\mu$m full thickness devices with both high and low resistivity epitaxial layers. 

Current limited latch-up states have been observed with a typical increase of 300~mA in the operating digital current. Damage similar to the one observed in the STAR environment, with permanent increase of operating digital current and sensor data corruption, has been reproduced only on thinned high-resistivity sensors. 
\begin{figure}[h!]
\centering
\includegraphics[width=.50\textwidth]{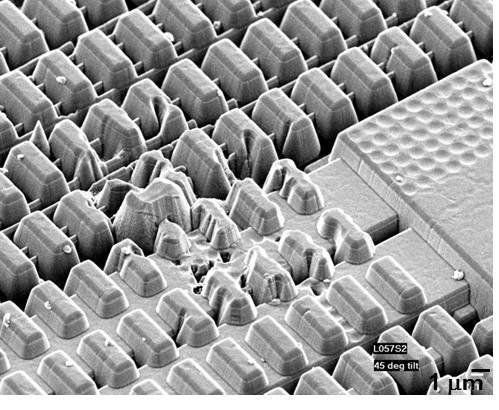}
\caption{\label{fig:SEM_damage}Scanning electronic microscope image of a damaged PXL sensor deconstructed through plasma etching technique (image taken at the \emph{Instrumentation Division laboratories - Brookhaven National Laboratory}). The metal layer appears to be melted.}
\end{figure}
A further analysis of the damaged sensors through infrared camera inspection, located the damage in specific structures of the digital section. The sensor substrate and sensitive epi-layer in the region of interest have been removed via plasma etching technique and examined through scanning electronic microscope at the BNL Instrumentation Division laboratories (see figure~\ref{fig:SEM_damage}), showing a modification of the metal layer, which appears to be melted.

In the following years of operations, an over-current protection threshold at 80~mA above the operating current ($\sim$1~A for the digital power) was applied to the running detector and limited significantly the onset of latch-up induced damage to 5 inner layer sensors per each RHIC Run in 2015 and 2016.
\begin{figure}[htbp]
\centering 
\includegraphics[width=.495\textwidth, trim=0 0 0.7cm 0]{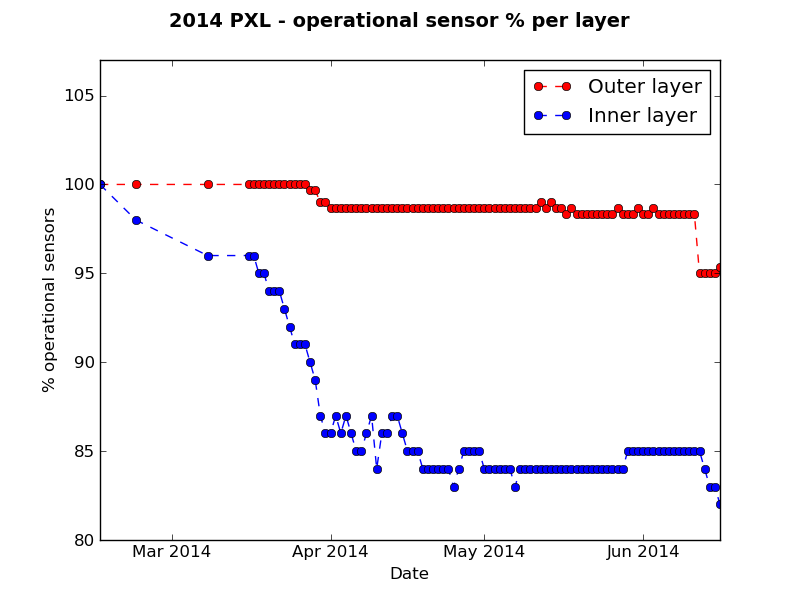}
\includegraphics[width=.495\textwidth, trim=0 0 0.7cm 0]{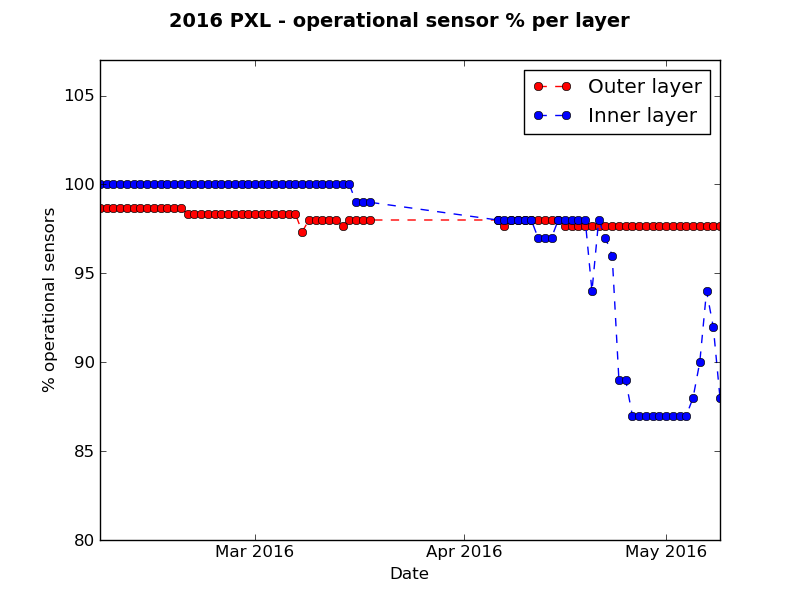}
\caption{\label{fig:damage}PXL detector active fraction evolution for the inner (\emph{blue}) and outer (\emph{red}) layers. During the 2014 Run (\emph{left}), most of the damage happened during the first weeks of Au+Au collisions data taking. The damage was then limited through operational methods. The drop observed at the end of the 2016 Run (\emph{right}) is due to one single damaged sensor that prevented the ladder from being operated in a stable state.}
\end{figure}

\paragraph{Firmware}
A significant loss of PXL efficiency was discovered in the 2015 Run data after the end of the operations, resulting in $\sim$40\% efficiency on single HFT track reconstruction with respect to the efficiency measured in 2014. The cause was identified in the PXL readout firmware version deployed at the beginning of 2015. As demonstrated through a post-run investigation based on sensor illumination with an external LED, an unsafe implementation of a timing constraint in the RDO board FPGA firmware caused the sensor frame reconstruction to fail: in each triggered PXL event, a fraction of reconstructed sensor data is not associated with the event itself. The extensive firmware tests carried out with pattern data and the performance of full detector calibrations before the firmware deployment were inadequate to spot this problem. In 2016 the PXL detector took data with the correct readout firmware version and a fast offline QA tool was put in place prior to the Run, allowing for fast detection of potential inefficiencies in the HFT data. 
A proper fix to the flawed firmware version was proposed and tested on the bench and in STAR at the end of the 2016 Run. Its efficiency matches now the efficiency measured in the 2014 and 2016 Runs. This new firmware version is available for future use.
The PXL experience also demonstrates how short-lived detector projects can suffer from the limited time available to tune readout firmware and reconstruction software.

\section{Conclusions}\label{sec:conclusions}
The STAR PXL detector collected more than 3 Billion minimum-bias Au+Au events at $\sqrt{s_{NN}}$~=~200~GeV, plus additional d+Au, p+Au and p+p samples, during the three-year (2014-2016) physics program at RHIC. The detector operations and subsequent physics data analysis demonstrate that the MAPS technology is an excellent choice for vertex detectors in the RHIC environment.\\
The PXL detector represents a breakthrough in the field, thanks to the achieved material budget (0.4\% $X_0$) and the novel mechanics design.
These parameters, combined with the \emph{\mbox{Ultimate-2}} sensor characteristics in terms of pixel pitch (20.7~$\mu$m) and thickness ($\sim$50~$\mu$m), allowed achieving a track pointing resolution of $\sim$46~$\mu$m for 750~MeV/c kaons. The PXL performance met the design requirements and enabled STAR to access the heavy flavor domain and to study the charmed hadron production at RHIC. \\
Continuous improvements in feature size, radiation hardness and readout speed will allow utilizing the MAPS technology in future particle physics applications.
Following the successful experience in STAR, the next-generation MAPS sensors will be used for the ALICE Inner Tracking System (ITS) Upgrade \cite{ALICE} at LHC and the CBM Micro-Vertex Detector (MVD) \cite{CBM} at FAIR, and have been proposed for the MAPS-based VerTeX detector (MVTX) of sPHENIX \cite{MVTX}, a next-generation nuclear physics experiment for multiscale studies of the strongly coupled quark-gluon plasma planned for the year 2022 and beyond at RHIC. Furthermore, Depleted MAPS \emph{(DMAPS)} sensors are being considered for the future proton-proton LHC experiments \cite{LHC}.

\section*{Acknowledgments}
This work was supported by the Director, Office of Science, Office of Nuclear Science of the U.S. Department of Energy under Contracts No. DE-AC02-05CH11231 and No. DE-SC0013391. We gratefully acknowledge the PICSEL group of IPHC Strasbourg (M. Winter et al.) for the development of the PXL detector sensors. We thank Robert Soja, William Christie and the STAR Technical and Detector Support Groups at Brookhaven National Laboratory for their help with installation, operations and maintenance of the PXL detector in STAR.

%
%
%

\end{document}